# Cosmic Microwave Background Polarization

Arthur Kosowsky[*]

*Department of Physics*
*Enrico Fermi Institute, The University of Chicago, Chicago, IL 60637-1433*
*and*
*NASA/Fermilab Astrophysics Center, Fermi National Accelerator Laboratory, Batavia, IL 60510-0500*
(December, 1994)

Polarization of the cosmic microwave background, though not yet detected, provides a source of information about cosmological parameters complementary to temperature fluctuations. This paper provides a complete theoretical treatment of polarization fluctuations. After a discussion of the physics of polarization, the Boltzmann equation governing the evolution of the photon density matrix is derived from quantum theory and applied to microwave background fluctuations, resulting in a complete set of transport equations for the Stokes parameters from both scalar and tensor metric perturbations. This approach is equivalent at lowest order in scattering kinematics to classical radiative transfer, and provides a general framework for treating the cosmological evolution of density matrices. The metric perturbations are treated in the physically appealing longitudinal gauge. Expressions for various temperature and polarization correlation functions are derived. Detection prospects and theoretical utility of microwave background polarization are briefly discussed.

## I. INTRODUCTION

Since the initial announcement by the COBE team of the detection of cosmic microwave background temperature anisotropies [1], a great deal of experimental activity has resulted in nearly a dozen more anisotropy detections on a wide range of angular scales [2,3]. Simultaneously, detailed numerical analysis has sharpened theoretical expectations for the temperature anisotropy and its dependence on a variety of cosmological parameters, primarily in the context of theories with initial adiabatic perturbations (of which Cold Dark Matter is a special case) [4–7] but also in cosmological defect models [8]. While much work remains to be done, the focus of microwave background research has shifted from simply detecting anisotropies to creating a detailed picture, both experimental and theoretical, of the anisotropies on all angular scale, and using this picture to constrain cosmological models [3].

The cosmological information in the microwave background is encoded not only in temperature fluctuations but also in its polarization. Since, as discussed below in Sec. II, the source term for generating polarization is fluctuations in radiation intensity, generally polarization fluctuations are expected to be somewhat smaller than temperature fluctuations. Numerical calculations have confirmed this rough expectation, giving polarization fluctuations no larger than 10% of the temperature anisotropies [9,10]. Greater experimental sensitivity is required to measure polarization than temperature fluctuations, and so far only upper limits have been established. However, polarization fluctuation measurements also have certain experimental advantages over temperature fluctuation measurements, making first detection of polarization within the next few years a reasonable possibility.

Currently the best polarization limit comes from the Saskatoon experiment [2], with a 95% confidence level upper limit of 25 $\mu$K between two orthogonal linear polarizations at angular scales of about a degree, corresponding to $9 \times 10^{-6}$ of the mean temperature. An earlier experiment mapped a large portion of the sky with a 7° beam to place upper limits of $6 \times 10^{-5}$ in linear polarization and $6 \times 10^{-4}$ in circular polarization for quadrupole and octupole variations [11], while a later measurement put limits of around $5 \times 10^{-5}$ on both linear and circular polarization at arcminute scales [12]. The COBE satellite made polarized measurements and can in principle achieve a limit of around $10^{-5}$ at angular scales from 7° to quadrupole, although doing so would require reanalysis of the entire data set [13]. While the sensitivity of the Saskatoon experiment is very good, it is designed primarily to measure temperature fluctuations. A new experiment planned for 1995 is optimized to measure polarization at 1° and 7° scales and aims for a sensitivity at or below $10^{-6}$ [14].

On the theory side, microwave background polarization has been discussed for many years [15–19]. Detailed calculations for CDM models have given linear polarization as large as 10% of the temperature anisotropy at medium angular scales, with a strong dependence on the ionization history of the universe [9,10]. Another recent calculation has mapped expected correlation patterns between temperature fluctuations and polarization [20].

---

[*]Current Address: Harvard/Smithsonian Center for Astrophysics, Mail Stop 51, 60 Garden Street, Cambridge, MA 02138; akosowsky@cfa.harvard.edu

The aim of this paper is a detailed investigation of the theory of microwave background polarization. In contrast to previous work employing classical radiative transfer theory [21], the evolution of polarization is derived from a photon description; this approach has previously been applied to temperature fluctuations in systematic investigations of second-order effects [6,22]. The usual classical Boltzmann equation, adequate for describing temperature fluctuations, must be generalized to a density matrix describing the photon polarization state. The formalism for this generalized Boltzmann equation was recently developed in the context of neutrino-flavor evolution [23]. Here the appropriate equation is derived beginning with the fundamental description of the relevant Compton scattering process; the techniques easily generalize to give a Boltzmann equation for any particular density matrix. The advantage of this approach is that it treats photons in a general manner, like other particle species described by a Boltzmann equation, and can easily be applied to other polarized distributions, e.g. electrons in a magnetic field. It also gives a systematic perturbative expansion in the relevant small quantities, and thus provides the framework for an investigation of all second-order polarization effects, which may be of particular interest in the case of early reionization for which the polarization contribution is the largest [24,25]. This paper also serves as a review which presents many derivations of relevant formulas which remain unpublished.

Section II provides an overview of the physics of polarization and its application to the microwave background. A simple calculation demonstrates that only quadrupolar variations in radiation intensity on a scatterer produce polarization. The Stokes parameters are defined and their connection to the photon density matrix made explicit. Section III derives the general formula for the time evolution of a density matrix in terms of an interaction Hamiltonian. This section is rather formal; the relevant result is Eq. (3.12). Section IV specializes this result to the evolution of the photon density matrix including Compton scattering. The calculation in this section is straightforward but long; the ultimate result is Eq. (4.24), which is equivalent to the usual classical equations of radiative transfer [21].

The particular application to a cosmological context begins in Section V, which derives the general relativistic Liouville equation for a perturbed Friedmann-Robertson-Walker spacetime. This collisionless part of the Boltzmann equation describes the photon geodesics in a homogeneous and isotropic universe with small scalar and tensor metric perturbations. The scalar perturbations are described with the physically appealing longitudinal gauge [26] instead of the more traditional synchronous gauge. Section VI gives the final evolution equations for the Stokes parameters describing the microwave background. The temperature and polarization fluctuations must be expressed in terms of their statistical properties for comparison with experimental results; Section VII derives expressions for the power spectra of various correlations and cross-correlations between temperature and polarization. Finally, the concluding section briefly discusses what detailed polarization measurements may eventually reveal about cosmology.

This paper employs natural units throughout, in which $\hbar = c = G = k_B = 1$; Section II uses gaussian units for electromagnetic quantities. The metric signature is $(+\ -\ -\ -)$. Spinor normalizations and other field theory conventions in Section IV conform to Mandl and Shaw [27].

## II. PHYSICS OF POLARIZATION

This section gives a qualitative overview of the physics of polarization in the context of the microwave background. We begin with a review of Stokes parameters, the conventional method for describing polarized light. Then we show how polarization can be generated by scattering; application to processes on the last scattering surface predict distinctive correlations between hot and cold spots and polarization direction. The equivalent polarization description in terms of the photon density matrix is then presented, with the connection to the conventional Stokes parameters made explicit.

### A. Review of Stokes Parameters

Polarized light is conventionally described in terms of the Stokes parameters, which are presented in any optics text [28]. Consider a nearly monochromatic plane electromagnetic wave propogating in the $z$-direction; nearly monochromatic here means that its frequency components are closely distributed around its mean frequency $\omega_0$. The components of the wave's electric field vector at a given point in space can be written as

$$E_x = a_x(t) \cos\left[\omega_0 t - \theta_x(t)\right], \quad E_y = a_y(t) \cos\left[\omega_0 t - \theta_y(t)\right]. \tag{2.1}$$

The requirement that the wave is nearly monochromatic guarantees that the amplitudes $a_x$ and $a_y$ and the phase angles $\theta_x$ and $\theta_y$ will be slowly varying functions of time relative to the inverse frequency of the wave. If some correlation exists between the two components in Eq. (2.1), then the wave is polarized.

The Stokes parameters are defined as the following time averages:

$$I \equiv \langle a_x^2 \rangle + \langle a_y^2 \rangle; \tag{2.2a}$$
$$Q \equiv \langle a_x^2 \rangle - \langle a_y^2 \rangle; \tag{2.2b}$$
$$U \equiv \langle 2a_x a_y \cos(\theta_x - \theta_y) \rangle; \tag{2.2c}$$
$$V \equiv \langle 2a_x a_y \sin(\theta_x - \theta_y) \rangle. \tag{2.2d}$$

The parameter $I$ gives the intensity of the radiation which is always positive. The other three parameters define the polarization state of the wave and can have either sign. Unpolarized radiation, or "natural light," is described by $Q = U = V = 0$. One important property of the Stokes parameters is that they are additive for incoherent superpositions of waves. The four parameters can be measured with a linear polarizer and a quarter-wave plate; the first three can be measured with only a linear polarizer. The $V$ parameter can also be measured as the intensity difference between left and right circular polarizations.

The parameters $I$ and $V$ are physical observables independent of the coordinate system, but $Q$ and $U$ depend on the orientation of the $x$ and $y$ axes. If a given wave is described by the parameters $Q$ and $U$ for a certain orientation of the coordinate system, then after a rotation of the $x - y$ plane through an angle $\phi$, the same wave is now described by the parameters

$$\begin{aligned} Q' &= Q\cos(2\phi) + U\sin(2\phi), \\ U' &= -Q\sin(2\phi) + U\cos(2\phi). \end{aligned} \tag{2.3}$$

From this transformation it is easy to see that the quantity $Q^2 + U^2$ is invariant under rotation of the axes, and the angle

$$\alpha \equiv \frac{1}{2}\tan^{-1}\frac{U}{Q} \tag{2.4}$$

transforms to $\alpha - \phi$ under a rotation by $\phi$ and thus defines a constant direction. The physically observable polarization vector **P** is here defined as orthogonal to the direction of wave propogation, having magnitude $(Q^2 + U^2)^{1/2}$ and polar angle $\alpha$. For a wave with linear polarization, the vector **P** lies along the constant orientation of the electric field. Note that since the definition (2.4) is degenerate for $\alpha$ and $\alpha + \pi$, only the orientation of **P** is defined and not the direction. We take the range of $\alpha$ to be $-\pi/2 < \alpha < \pi/2$ with the sign of $\alpha$ the same as the sign of $U$. While the radiation transport equations below are most conveniently formulated in terms of the Stokes parameters, the physical interpretation of a polarization pattern is clearest in terms of the observables $I$, $V$, and **P**.

### B. Polarization and the Last Scattering Surface

In the early universe, at redshifts greater than about $z \approx 1100$, the baryons, electrons, and photons comprise a tightly coupled fluid. Small metric perturbations induce bulk velocities of the fluid, and the resulting anisotropies in the photon distribution will induce polarization when the photons scatter off charged particles. After recombination, the photons freely propagate along geodesics, and any polarization produced before recombination will remain fixed. A sufficiently early reionization can of course generate further polarization.

A simple idealization of the last scattering surface elucidates the process of polarization generation. Consider initially unpolarized light which undergoes Thomson scattering at a given point and is then viewed by an observer. If the intensity of the light incident on the scattering point is uniform in every direction, then obviously no polarization can result; however, if the incident intensity varies with direction then polarization can be generated. Choose the $z$-axis to lie in the direction of the outgoing light, which is described by the Stokes parameters $I$, $Q$, $U$, and $V$; represent the light incident on the scattering point by the intensity $I'(\theta, \phi)$. Define the polarization vectors for the outgoing beam of light so that $\hat{\epsilon}_x$ is perpendicular to the scattering plane and $\hat{\epsilon}_y$ is in the scattering plane, and likewise with the incoming polarization vectors $\hat{\epsilon}'_x$ and $\hat{\epsilon}'_y$ (see Fig. 1). Also, instead of dealing with $I$ and $Q$, it is convenient to describe the scattering process in terms of $I_x = (I + Q)/2$ and $I_y = (I - Q)/2$. The Thomson scattering cross-section for an incident wave with linear polarization $\hat{\epsilon}'$ into a scattered wave with linear polarization $\hat{\epsilon}$ is given by

$$\frac{d\sigma}{d\Omega} = \frac{3\sigma_T}{8\pi}|\hat{\epsilon}' \cdot \hat{\epsilon}|^2 \tag{2.5}$$

where $\sigma_T$ is the total Thomson cross section. The incoming wave is unpolarized by assumption, and thus satisfies $I'_x = I'_y \equiv I'/2$. The scattered intensities are

$$I_x = \frac{3\sigma_T}{8\pi}\left[I'_x(\hat{\epsilon}'_x \cdot \hat{\epsilon}_x)^2 + I'_y(\hat{\epsilon}'_y \cdot \hat{\epsilon}_x)^2\right] = \frac{3\sigma_T}{16\pi}I', \tag{2.6a}$$

$$I_y = \frac{3\sigma_T}{8\pi}\left[I'_x(\hat{\epsilon}'_x \cdot \hat{\epsilon}_y)^2 + I'_y(\hat{\epsilon}'_y \cdot \hat{\epsilon}_y)^2\right] = \frac{3\sigma_T}{16\pi}I'\cos^2\theta. \tag{2.6b}$$

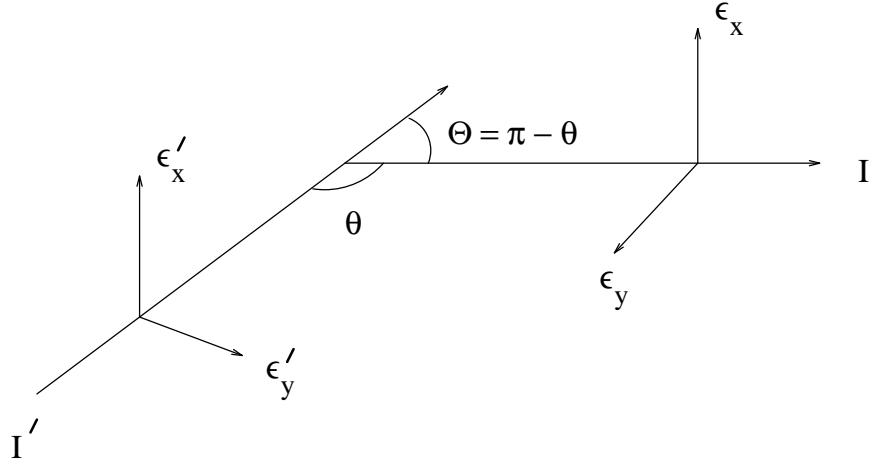

FIG. 1. Definition of vectors and angles for Thomson scattering of a light beam or photon.

Thus the scattered wave has the Stokes parameters

$$I = I_x + I_y = \frac{3\sigma_T}{16\pi}I'(1+\cos^2\theta), \tag{2.7a}$$

$$Q = I_x - I_y = \frac{3\sigma_T}{16\pi}I'\sin^2\theta. \tag{2.7b}$$

This calculation gives no information about the $U$ or $V$ parameters. As will be shown later, the $V$ parameter remains zero after scattering and will not be considered further [21]. The $U$ parameter can be determined by using Eq. (2.3). Simply rotate the outgoing basis vectors in the above calculation by $\pi/4$ and recalculate $Q$, which will be equal to $U$ in the original coordinate system. The result is $U = 0$. These results can alternatively be obtained from the physical description of the polarization state in Rayleigh scattering [21]. Note that Eq. (2.7b) gives the well-known result that sunlight from the horizon at midday is linearly polarized parallel to the horizon.

The total scattered intensities are determined by integrating over all incoming intensities. Note that the outgoing $U$ and $Q$ flux from a given incoming direction must always be rotated into a common coordinate system, using Eq. (2.3). The result is

$$I = \frac{3\sigma_T}{16\pi}\int d\Omega\,(1+\cos^2\theta)I'(\theta,\phi), \tag{2.8a}$$

$$Q = \frac{3\sigma_T}{16\pi}\int d\Omega\,\sin^2\theta\cos(2\phi)I'(\theta,\phi), \tag{2.8b}$$

$$U = -\frac{3\sigma_T}{16\pi}\int d\Omega\,\sin^2\theta\sin(2\phi)I'(\theta,\phi). \tag{2.8c}$$

The outgoing polarization state depends only on the intensity distribution of the unpolarized incident radiation. Expanding the incident intensity in spherical harmonics,

$$I'(\theta,\phi) = \sum_{lm} a_{lm}Y_{lm}(\theta,\phi), \tag{2.9}$$

leads to the following expressions for the outgoing Stokes parameters:

$$I = \frac{3\sigma_T}{16\pi}\left[\frac{8}{3}\sqrt{\pi}\,a_{00} + \frac{4}{3}\sqrt{\frac{\pi}{5}}\,a_{20}\right], \tag{2.10a}$$

$$Q = \frac{3\sigma_T}{4\pi}\sqrt{\frac{2\pi}{15}}\,\mathrm{Re}\,a_{22}, \tag{2.10b}$$

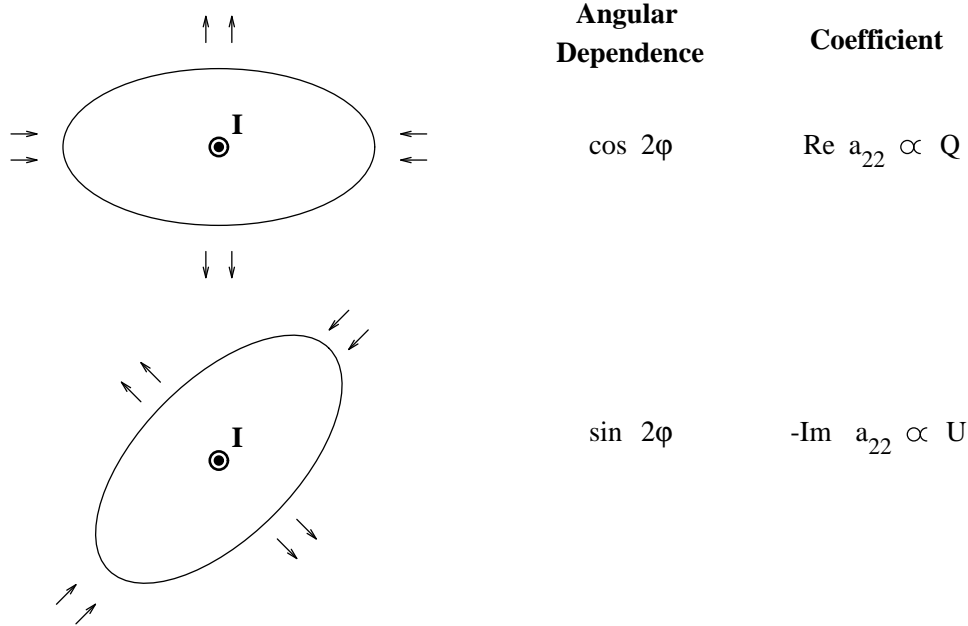

FIG. 2. The quadrupolar components of the incident intensity distribution. Any orientation of a quadrupolar distribution can be written as the sum of these two distributions. The small arrows indicate the corresponding fluid velocity in a tightly coupled fluid.

$$U = -\frac{3\sigma_T}{4\pi}\sqrt{\frac{2\pi}{15}}\,\mathrm{Im}\,a_{22}. \qquad (2.10c)$$

Thus scattering generates polarization from initially unpolarized radiation if the radiation intensity at a given point as a function of direction has a non-zero component of $Y_{22}$.

This particular form for the source of polarization leads to a correlation of the direction of the polarization vector **P** with hot and cold spots on the cosmic microwave sky [20]. Consider a given region on the last scattering surface with a spherical mass overdensity; the electron-photon fluid will have a bulk velocity towards the center of the overdense region with a velocity gradient away from the center (material further from the center will be falling inwards more quickly). In the frame of some particular scattering point away from the center, the fluid velocity in towards the point is greater along the radial direction than perpendicular to it, resulting in a quadrupolar radiation intensity variation with the largest intensity along the radial direction (see Figure 2). Choose an observation direction at a right angle to the radial direction and take this direction to be the polar axis. Then the radiation intensity at the scattering point will have a component proportional to $\cos(2\phi - 2\beta)$ with a positive coefficient, where $\beta$ is the radial direction. The scattered $Q$ intensity is proportional to the $\cos(2\phi)$ dependence of the incident intensity and the scattered $U$ intensity is proportional to the $\sin(2\phi)$ piece, by Eqs. (2.9) and (2.10). Thus $\alpha$, the direction of **P** in Eq. (2.4), lies along $\beta$, the radial direction. For the opposite situation, that of a mass underdensity, all the velocities change sign, so both $Q$ and $U$ change sign and the direction of **P** changes by $\pi/2$. When the dominant contribution to the temperature fluctuations is a gravitational potential difference (Sachs-Wolfe effect [29]), a mass overdensity corresponds to a cold spot in the microwave background; in this case cold spots will have radially correlated polarization and hot spots tangentially correlated polarization, in agreement with the result of Ref. [20]. For adiabatic acoustic oscillations, the density and velocity perturbations are out of phase so no specific correlation results.

### C. Photon Description

The Stokes parameters can be defined equivalently in terms of a quantum-mechanical description. The polarization state space of a photon is spanned by a pair of basis vectors, which we take to be the orthogonal linear polarizations $|\epsilon_1\rangle$ and $|\epsilon_2\rangle$. For a photon propogating in the z-direction, the basis states $|\epsilon_1\rangle$ and $|\epsilon_2\rangle$ are oriented like the $x$ and $y$ axes, respectively. An arbitrary state is given by

$$|\epsilon\rangle = a_1 e^{i\theta_1}|\epsilon_1\rangle + a_2 e^{i\theta_2}|\epsilon_2\rangle. \qquad (2.11)$$

The quantum-mechanical operators in the linear basis corresponding to each Stokes parameter are given by

$$\hat{I} = |\epsilon_1\rangle\langle\epsilon_1| + |\epsilon_2\rangle\langle\epsilon_2|; \tag{2.12a}$$

$$\hat{Q} = |\epsilon_1\rangle\langle\epsilon_1| - |\epsilon_2\rangle\langle\epsilon_2|; \tag{2.12b}$$

$$\hat{U} = |\epsilon_1\rangle\langle\epsilon_2| + |\epsilon_2\rangle\langle\epsilon_1|; \tag{2.12c}$$

$$\hat{V} = i|\epsilon_2\rangle\langle\epsilon_1| - i|\epsilon_1\rangle\langle\epsilon_2|. \tag{2.12d}$$

The single-particle state expectation values of these operators reproduce the definitions (2.2). For photons in a general mixed state defined by a density matrix $\rho$, the expectation value for the $I$ Stokes parameter is given by

$$\langle I \rangle = \operatorname{tr}\rho\hat{I} = \operatorname{tr}\begin{pmatrix} \rho_{11} & \rho_{12} \\ \rho_{21} & \rho_{22} \end{pmatrix}\begin{pmatrix} 1 & 0 \\ 0 & 1 \end{pmatrix} = \rho_{11} + \rho_{22} \tag{2.13}$$

and similarly for the other three parameters. These relations thus give the density matrix in the linear polarization basis in terms of the Stokes parameters as

$$\rho = \frac{1}{2}\begin{pmatrix} I+Q & U-iV \\ U+iV & I-Q \end{pmatrix} \tag{2.14}$$

$$= \frac{1}{2}(I\mathbb{1} + Q\sigma_3 + U\sigma_1 + V\sigma_2) \tag{2.15}$$

where $\mathbb{1}$ is the identity matrix and $\sigma_i$ are the Pauli spin matrices. Thus the density matrix for a system of photons contains the same information as the four Stokes parameters, and the time evolution of the density matrix gives the time evolution of the system's polarization.

## III. EVOLUTION EQUATION FOR THE NUMBER OPERATOR

This section considers the quantum number operator for a system of particles and derives its evolution equation, including local particle interactions. Taking the expectation value of the operator equation gives the Boltzmann equation for the system's density matrix, which is a generalization of the usual classical Boltzmann equation for particle occupation numbers (the diagonal elements of the density matrix). The derivation here applies techniques previously developed in the context of neutrino mixing [23,30].

We adopt second-quantized formalism with creation and annihilation operators for photons and electrons obeying the canonical commutation relations

$$[a_s(p), a_{s'}^\dagger(p')] = (2\pi)^3 2p^0 \delta^3(\mathbf{p}-\mathbf{p}')\delta_{ss'}, \tag{3.1}$$

$$\{b_r(q), b_{r'}^\dagger(q')\} = (2\pi)^3 \frac{q^0}{m}\delta^3(\mathbf{q}-\mathbf{q}')\delta_{rr'}, \tag{3.2}$$

where $s$ labels the photon polarization and $r$ labels the electron spin; bold momentum variables represent three-momenta while plain momentum variables represent four-momenta. The density operator describing a system of photons is given by

$$\hat{\rho} = \int \frac{d^3\mathbf{p}'}{(2\pi)^3}\rho_{ij}(\mathbf{p}')a_i^\dagger(\mathbf{p}')a_j(\mathbf{p}'), \tag{3.3}$$

where $\rho_{ij}$ is the density matrix. The particular operator for which we want the equation of motion is the photon number operator

$$\mathcal{D}_{ij}(\mathbf{k}) \equiv a_i^\dagger(\mathbf{k})a_j(\mathbf{k}). \tag{3.4}$$

The expectation value of $\mathcal{D}$ is proportional to the density matrix, as seen by direct calculation:

$$\langle\mathcal{D}_{ij}(\mathbf{k})\rangle = \operatorname{tr}[\hat{\rho}\mathcal{D}_{ij}(\mathbf{k})] = \int \frac{d^3\mathbf{p}}{(2\pi)^3}\langle\mathbf{p}|\hat{\rho}\mathcal{D}_{ij}(\mathbf{k})|\mathbf{p}\rangle = (2\pi)^3\delta(0)2k^0\rho_{ij}(\mathbf{k}). \tag{3.5}$$

The last equality results from repeated application of the commutation relation Eq. (3.2); the infinite delta function results from the infinite quantization volume necessary with continuous momentum variables, and cancels out of all physical results.

The time evolution of the operator $\mathcal{D}_{ij}$, considered in the Heisenberg picture, is

$$\frac{d}{dt}\mathcal{D}_{ij} = i\,[H, \mathcal{D}_{ij}] \tag{3.6}$$

where $H$ is the full Hamiltonian. We write the Hamiltonian as a sum of the free field piece plus an interaction term:

$$H = H_0 + H_I \tag{3.7}$$

where the interaction piece is a functional of the full fields in the problem. Our goal is to express the right side of Eq. (3.6) as a perturbation series in the interaction Hamiltonian $H_I$. We make the usual assumption of scattering theory that in a given interaction the fields begin as free fields and end as other free fields, and the interactions are isolated from each other. Consider the evolution of an operator through a single interaction beginning at $t = 0$: before this time, the fields can be taken as free to a good approximation; at $t = 0$ the interaction Hamiltonian begins to turn on, and the interaction finishes at some later time, after which the fields can be taken as free once again. Then the time dependence of an arbitrary operator $\xi$ to first order in the interaction Hamiltonian can be expressed as [23]

$$\xi(t) = \xi^0(t) + i\int_0^t dt'\,[H_I^0(t-t'), \xi^0(t)], \tag{3.8}$$

where $\xi^0(t)$ is the free-field operator with initial condition $\xi^0(0) = \xi(0)$, and $H_I^0$ is the interaction Hamiltonian as a functional of the free fields.

Equation (3.8) can be proven as follows. The time derivative of both sides gives

$$\frac{d}{dt}\xi(t) = \frac{d}{dt}\xi^0(t) + i\,[H_I^0(0), \xi^0(t)] + i\int_0^t dt'\,\frac{d}{dt}\,[H_i^0(t-t'), \xi^0(t)]. \tag{3.9}$$

The time derivatives in the first and third term on the right side can be replaced by commutators with $H(t)$ using the Heisenberg equation. But these two terms depend only on free fields which are evolved with the free Hamiltonian $H_0(t) = H_0(0)$. Equation (3.9) becomes

$$\frac{d}{dt}\xi(t) = i\,[H_I^0(0), \xi^0(t)] + i\,[H_0(0), \xi(t)], \tag{3.10}$$

and so to first order in $H_I$ this just gives the Heisenberg equation for the operator $\xi$.

Now we can express the time evolution of $\mathcal{D}_{ij}$ in terms of free field operators. Applying Eq. (3.8) to the commutator on the right side of Eq. (3.6) gives

$$\frac{d}{dt}\mathcal{D}_{ij}(\mathbf{k}) = i\,[H_I^0(t), \mathcal{D}_{ij}^0(\mathbf{k})] - \int_0^t dt'\,[H_I^0(t-t'), [H_I^0(t), \mathcal{D}_{ij}^0(\mathbf{k})]]. \tag{3.11}$$

The integral on the right side can be cast in a more practical form by making the following physical assumption: the duration of each collision (the time interval over which the interaction Hamiltonian is non-negligible, on the order of the inverse energy transfer) is small compared to the time scale for variation of the density matrix (on the order of the inverse collision frequency). The collision process relevant to the microwave background is Compton scattering off electrons, and for the cosmological epoch of interest, the electron density is always low enough for this condition to be easily satisfied. Then the time step $t$ in Eq. (3.11) can be chosen large compared with a single collision and small compared to the time scale for density matrix evolution. After extending the time integral to infinity and taking the expectation value of both sides, we find [23]

$$(2\pi)^3\delta(0)2k^0\frac{d}{dt}\rho_{ij}(\mathbf{k}) = i\,\langle[H_I^0(t), \mathcal{D}_{ij}^0(\mathbf{k})]\rangle - \frac{1}{2}\int_{-\infty}^{\infty} dt\,\langle[H_I^0(t), [H_I^0(0), \mathcal{D}_{ij}^0(\mathbf{k})]]\rangle. \tag{3.12}$$

Here the integral from zero to infinity has been replaced by an integral over all time; the difference is a principle part integral which is second-order in the interaction Hamiltonian and thus ignored.

Equation (3.12) is the Boltzmann equation for the density matrix $\rho_{ij}$. The first term on the right side is a forward scattering term which is responsible for the MSW effect in a neutrino ensemble [30]; for photons this term is zero, as will be shown below. The second term on the right side is the usual collision term. The time integral over the free field time dependence enforces energy conservation in each collision. The interaction Hamiltonian will in most cases depend on background fields; for example, in the case of the microwave background the Compton scattering collisions are essentially four-point interactions quadratic in both the photon field and the electron field. In principle, a second coupled equation for the electron density matrix must be solved simultaneously. However, in many physical situations, the background fields may be assumed to have a fixed distribution, generally thermal. In the early universe, the electrons maintain a thermal distribution to a very high approximation and the evolution of their density matrix becomes trivial.

The derivation of Eq. (3.12) has been completely general. In the appropriate limit the classical equations of radiative transfer are reproduced; the advantage to the current approach is that it provides the same formal framework for treating polarized photons as for treating neutrinos or any other particle species governed by the Boltzmann equation. It also gives a systematic method for analyzing all higher-order effects. This Boltzmann equation has previously been applied to neutrinos interacting through both charged and neutral current processes in supernovae [31].

## IV. APPLICATION TO COMPTON SCATTERING OF PHOTONS

In principle, the complete evolution of the cosmic microwave background is determined by Eq. (3.12), generalized slightly to include spatial dependence of all quantities. In this work this space dependence will simply be put in by hand when taking expectation values and assumed implicitly; more formally it can be included through Wigner functions, describing a joint space-momentum distribution [32]. All that remains to be done is substitution of the correct interaction Hamiltonian and simplification of the right side. General relativistic terms emerge from the total time derivative on the left side; these will be treated in detail in Sec. V.

Microwave background photons interact with all charged particles. However, the rate of scattering varies with the mass of the charged particle as the inverse mass squared; thus it is an excellent approximation to consider only Compton scattering off electrons and ignore baryons. This section proceeds with evaluation of the right side of Eq. (3.12) for Compton scattering.

### A. Interaction Hamiltonian

The interaction Hamiltonian density for the fundamental three-point interaction of QED is given by [27]

$$\mathcal{H}_{\rm QED}(x) = -e : \bar{\psi}(x) \slashed{A}(x) \psi(x) : \tag{4.1}$$

where $\psi$ is the electron field operator, $A^\mu$ is the photon field operator, a slash indicates contraction with $\gamma_\mu$, and the colons signify normal ordering of the operator product. The interaction Hamiltonian is the density integrated over all of space:

$$H_{\rm QED}(t) = \int d^3\mathbf{x} \mathcal{H}_{\rm QED}(x). \tag{4.2}$$

The scattering matrix describing all scattering processes in QED is given in terms of the interaction Hamiltonian by

$$S = \sum_{n=0}^{\infty} S^{(n)} \equiv \sum_{n=0}^{\infty} \frac{(-i)^n}{n!} \int d^4x_1 \ldots d^4x_n \mathrm{T}\{\mathcal{H}_{\rm QED}(x_1) \ldots \mathcal{H}_{\rm QED}(x_n)\} \tag{4.3}$$

where T signifies a time-ordered product. The nth term in the series represents all scattering processes with $n$ interaction vertices. Compton scattering is thus contained in the $n = 2$ term of the scattering matrix. Comparing the $n = 2$ term with the $n = 1$ term gives the interaction Hamiltonian for second-order scattering processes:

$$\begin{aligned} S^{(2)} &= -\frac{1}{2} \int_{-\infty}^{\infty} dt \int_{-\infty}^{\infty} dt' \mathrm{T}\{H_{\rm QED}(t) H_{\rm QED}(t')\} \\ &\equiv -i \int_{-\infty}^{\infty} dt H^{(2)}(t). \end{aligned} \tag{4.4}$$

Using Wick's theorem to simplify the time-ordered product and denoting the piece of $H^{(2)}$ describing Compton scattering by $H_I$ yields

$$H_I(t) = e^2 \int_{-\infty}^{\infty} dt' \int d^3\mathbf{x}' \bar{\psi}^-(x) \gamma^\alpha S_F(x-x') \gamma^\beta \left[ A_\alpha^-(x) A_\beta^+(x') + A_\beta^-(x') A_\alpha^+(x) \right] \psi^+(x') \tag{4.5}$$

where $S_F$ is the Feynman propagator for the electron, and $\psi^+$ ($\bar{\psi}^-$) and $A^+$ ($A^-$) are linear in absorption (creation) operators of electrons and photons respectively. Fourier transforms of the fields and propagator are defined using the following conventions:

$$A^\mu(x) = \int \frac{d^3\mathbf{k}}{(2\pi)^3 2k^0} \sum_s \left[ a_s(k) \varepsilon_s^\mu(k) e^{-ik\cdot x} + a_s^\dagger(k) \varepsilon_s^{\mu*}(k) e^{ik\cdot x} \right], \tag{4.6a}$$

$$\bar{\psi}^-(x) = \int \frac{d^3\mathbf{k}}{(2\pi)^3} \frac{m}{k^0} \sum_r b_r^\dagger(k) \bar{u}_r(k) e^{ik\cdot x}, \tag{4.6b}$$

$$\psi^+(x) = \int \frac{d^3\mathbf{k}}{(2\pi)^3} \frac{m}{k^0} \sum_r b_r(k) u_r(k) e^{-ik\cdot x}, \tag{4.6c}$$

$$S_F(x) = \int \frac{d^4k}{(2\pi)^4} \frac{\slashed{k} + m}{k^2 - m^2 + i0} e^{-ik\cdot x}, \tag{4.6d}$$

where $u_r(k)$ is a spinor solution to the Dirac equation with spin index $r = 1, 2$ and $\varepsilon_s^\mu(k)$ are photon polarization four-vectors, chosen to be real, with index $s = 1, 2$ labeling the physical transverse polarizations of the photon. The summation convention over repeated spin and polarization indices is always implied. The Fourier-space interaction Hamiltonian is obtained by substituting Eqs. (4.6) into Eq. (4.5). The distributional identity

$$\int d^4x \, e^{ik\cdot x} = (2\pi)^4 \delta^4(k) \tag{4.7}$$

allows trivial integration over the four-momentum of the electron propagator. The resulting interaction Hamiltonian is

$$H_I^0(t) = \int d\mathbf{q}\, d\mathbf{q}'\, d\mathbf{p}\, d\mathbf{p}' (2\pi)^3 \delta^3(\mathbf{q}' + \mathbf{p}' - \mathbf{q} - \mathbf{p}) \exp\left[ it(q'^0 + p'^0 - q^0 - p^0) \right]$$
$$\times \left[ b_{r'}^\dagger(q') a_{s'}^\dagger(p') (\mathcal{M}_1 + \mathcal{M}_2) a_s(p) b_r(q) \right], \tag{4.8}$$

$$\mathcal{M} \equiv \mathcal{M}_1 + \mathcal{M}_2, \tag{4.9a}$$

$$\mathcal{M}_1(q'r', p's', qr, ps) \equiv e^2 \frac{\bar{u}_{r'}(q') \slashed{\varepsilon}_{s'}(p') (\slashed{p} + \slashed{q} + m) \slashed{\varepsilon}_s(p) u_r(q)}{2(p\cdot q)}, \tag{4.9b}$$

$$\mathcal{M}_2(q'r', p's', qr, ps) \equiv -e^2 \frac{\bar{u}_{r'}(q') \slashed{\varepsilon}_{s'}(p') (\slashed{q} - \slashed{p}' + m) \slashed{\varepsilon}_s(p) u_r(q)}{2(p'\cdot q)}, \tag{4.9c}$$

with the abbreviations

$$d\mathbf{q} \equiv \frac{d^3\mathbf{q}}{(2\pi)^3} \frac{m}{q^0}, \qquad d\mathbf{p} \equiv \frac{d^3\mathbf{p}}{(2\pi)^3 2p^0} \tag{4.10}$$

for electrons and photons respectively. All of the operators in Eq. (4.8) are free-field operators, so this is the proper expression to substitute into the left side of Eq. (3.12).

## B. Forward Scattering Term

We now proceed to evaluate the first term on the left side of Eq. (3.12). First we display operator expectation values needed here and in the following subsection, using operator definitions and the commutation relations, Eq. (3.2):

$$\langle a_1 a_2 \cdots b_1 b_2 \cdots \rangle = \langle a_1 a_2 \cdots \rangle \langle b_1 b_2 \cdots \rangle \tag{4.11a}$$

$$\langle a_m^\dagger(p') a_n(p) \rangle = 2p^0 (2\pi)^3 \delta^3(\mathbf{p} - \mathbf{p}') \rho_{mn}(\mathbf{p}) \tag{4.11b}$$

$$\langle b_m^\dagger(q') b_n(q) \rangle = \frac{q^0}{m} (2\pi)^3 \delta^3(\mathbf{q} - \mathbf{q}') \delta_{mn} \frac{1}{2} n_e(\mathbf{q}) \tag{4.11c}$$

$$\langle b_{r_1'}^\dagger(q_1') b_{r_1}(q_1) b_{r_2'}^\dagger(q_2') b_{r_2}(q_2) \rangle = \frac{q_1^0 q_2^0}{m^2} (2\pi)^6 \delta^3(\mathbf{q}_1 - \mathbf{q}_1') \delta^3(\mathbf{q}_2 - \mathbf{q}_2') \delta_{r_1 r_1'} \delta_{r_2 r_2'} \frac{1}{4} n_e(\mathbf{q}_1) n_e(\mathbf{q}_2)$$
$$+ \frac{q_1^0 q_2^0}{m^2} (2\pi)^6 \delta^3(\mathbf{q}_1 - \mathbf{q}_2') \delta^3(\mathbf{q}_2 - \mathbf{q}_1') \delta_{r_1 r_2'} \delta_{r_2 r_1'} \frac{1}{2} n_e(\mathbf{q}_2) \left[ 1 - \frac{1}{2} n_e(\mathbf{q}_1) \right] \tag{4.11d}$$

$$\langle a_{s_1'}^\dagger(p_1') a_{s_1}(p_1) a_{s_2'}^\dagger(p_2') a_{s_2}(p_2) \rangle = 4 p_1^0 p_2^0 (2\pi)^6 \delta^3(\mathbf{p}_1 - \mathbf{p}_1') \delta^3(\mathbf{p}_2 - \mathbf{p}_2') \rho_{s_1 s_1'}(\mathbf{p}_1) \rho_{s_2 s_2'}(\mathbf{p}_2)$$
$$+ 4 p_1^0 p_2^0 (2\pi)^6 \delta^3(\mathbf{p}_1 - \mathbf{p}_2') \delta^3(\mathbf{p}_2 - \mathbf{p}_1') \rho_{s_1' s_2}(\mathbf{p}_2) \left[ \delta_{s_1 s_2'} + \rho_{s_1 s_2'}(\mathbf{p}_1) \right] . \tag{4.11e}$$

The last relationship neglects the correlation term between all four operators when $p_1' = p_1 = p_2' = p_2$. The expectation values for electron operators assumes a particular form for the electron density matrix appropriate to thermal equilibrium, with equal populations in each spin state and no correlations between the states; $n_e(q)$ represents the number density of electrons of momentum $\mathbf{q}$ per unit volume. This assumed form for the electron density matrix will not be valid if substantial magnetic fields are present.

Using the definitions (3.4) and (4.8) and the commutation relations (3.2) the commutator in the forward scattering term becomes

$$\left[ H_I^0(0), \mathcal{D}_{ij}^0(\mathbf{k}) \right] = \int d\mathbf{q}\, d\mathbf{q}'\, d\mathbf{p}\, d\mathbf{p}' (2\pi)^3 \delta^3(\mathbf{q}' + \mathbf{p}' - \mathbf{q} - \mathbf{p}) \left( \mathcal{M}_1 + \mathcal{M}_2 \right)$$
$$\times \left[ b_{r'}^\dagger(q') b_r(q) a_{s'}^\dagger(p') a_j(k) 2 p^0 (2\pi)^3 \delta_{is} \delta^3(\mathbf{p} - \mathbf{k}) \right.$$
$$\left. - b_{r'}^\dagger(q') b_r(q) a_i^\dagger(k) a_s(p) 2 p'^0 (2\pi)^3 \delta_{js'} \delta^3(\mathbf{p}' - \mathbf{k}) \right]. \tag{4.12}$$

On using the above expectation values, it follows that

$$i \left\langle \left[ H_I^0(0), \mathcal{D}_{ij}^0(\mathbf{k}) \right] \right\rangle = \frac{i e^2}{4} \int d\mathbf{q}\, \frac{n_e(q)}{k \cdot q} \left( \delta_{is} \rho_{s'j}(\mathbf{k}) - \delta_{js'} \rho_{is}(\mathbf{k}) \right)$$
$$\times \bar{u}_r(q) \left[ \not{\varepsilon}_{s'}(k)(\not{k} + \not{q} + m) \not{\varepsilon}_s(k) - \not{\varepsilon}_s(k)(\not{q} - \not{k} + m) \not{\varepsilon}_{s'}(k) \right] u_r(q), \tag{4.13}$$

where the integrals have been performed with the delta functions. All of the terms involving $\not{k}$ cancel out on using the gamma-matrix identity $\not{A}\not{B} = 2 A \cdot B - \not{B}\not{A}$ and the polarization vector properties $k \cdot \varepsilon_i(k) = 0$ and $\varepsilon_i \cdot \varepsilon_j = -\delta_{ij}$. For the remaining terms we use the identity

$$\bar{u}_r(q) \not{\varepsilon}_s(\not{q} + m) \not{\varepsilon}_{s'} u_r(q) = \bar{u}_r(q)(2 q \cdot \varepsilon_s - \not{q}\not{\varepsilon}_s + m\not{\varepsilon}_s) \not{\varepsilon}_{s'} u_r(q)$$
$$= 2 q \cdot \varepsilon_s\, \bar{u}_r(q) \not{\varepsilon}_{s'} u_r(q)$$
$$= \frac{2}{m} (q \cdot \varepsilon_s)(q \cdot \varepsilon_{s'})$$
$$= \bar{u}_r(q) \not{\varepsilon}_{s'}(\not{q} + m) \not{\varepsilon}_s u_r(q), \tag{4.14}$$

where the second equality follows from the Dirac equation and the third equality uses the Gordon identity. Thus we have

$$i \left\langle \left[ H_I^0(0), \mathcal{D}_{ij}^0(\mathbf{k}) \right] \right\rangle = 0 \tag{4.15}$$

and the forward scattering term does not contribute to the photon density matrix evolution.

### C. Scattering Term

The scattering term is considerably more cumbersome to evaluate, being quadratic in the interaction Hamiltonian. After substituting the expressions for $H_I$ and $\mathcal{D}_{ij}$ and taking the expectation value, the scattering term reads

$$\frac{1}{2}\int_{-\infty}^{\infty} dt \langle [H_I^0(t), [H_I^0(0), \mathcal{D}_{ij}^0(\mathbf{k})]]\rangle = \int d\mathbf{q}_1 d\mathbf{q}_1' d\mathbf{p}_1 d\mathbf{p}_1' d\mathbf{q}_2 d\mathbf{q}_2' d\mathbf{p}_2 d\mathbf{p}_2'$$

$$\times (2\pi)^7 \delta^3(\mathbf{q}_1' + \mathbf{p}_1' - \mathbf{q}_1 - \mathbf{p}_1)\delta^4(q_2' + p_2' - q_2 - p_2)\mathcal{M}(1)\mathcal{M}(2)$$

$$\times \Bigg\{ p_1^0(2\pi)^3 \delta_{is_1}\delta^3(\mathbf{p}_1 - \mathbf{k}) \left\langle b_{r_2'}^\dagger(q_2') b_{r_2}(q_2) b_{r_1'}^\dagger(q_1') b_{r_1}(q_1)\right\rangle \left\langle a_{s_2'}^\dagger(p_2') a_{s_2}(p_2) a_{s_1'}^\dagger(p_1') a_j(k)\right\rangle$$

$$- p_1^0(2\pi)^3 \delta_{is_1}\delta^3(\mathbf{p}_1 - \mathbf{k}) \left\langle b_{r_1'}^\dagger(q_1') b_{r_1}(q_1) b_{r_2'}^\dagger(q_2') b_{r_2}(q_2)\right\rangle \left\langle a_{s_1'}^\dagger(p_1') a_j(k) a_{s_2'}^\dagger(p_2') a_{s_2}(p_2)\right\rangle$$

$$- p_1'^0(2\pi)^3 \delta_{js_1'}\delta^3(\mathbf{p}_1' - \mathbf{k}) \left\langle b_{r_2'}^\dagger(q_2') b_{r_2}(q_2) b_{r_1'}^\dagger(q_1') b_{r_1}(q_1)\right\rangle \left\langle a_{s_2'}^\dagger(p_2') a_{s_2}(p_2) a_i^\dagger(k) a_{s_1}(p_1)\right\rangle$$

$$+ p_1'^0(2\pi)^3 \delta_{js_1'}\delta^3(\mathbf{p}_1' - \mathbf{k}) \left\langle b_{r_1'}^\dagger(q_1') b_{r_1}(q_1) b_{r_2'}^\dagger(q_2') b_{r_2}(q_2)\right\rangle \left\langle a_i^\dagger(k) a_{s_1}(p_1) a_{s_2'}^\dagger(p_2') a_{s_2}(p_2)\right\rangle \Bigg\}. \quad (4.16)$$

The energy delta function comes from the time integral on using Eq. (4.7). The arguments of the matrix element indicates the subscript to be attached to all dependent variables in Eq. (4.9), and of course summation over all spin and polarization indices is implied.

Substitution of the expectation values Eq. (4.11) into the above expression and performing the integrals over $\mathbf{q}_2'$, $\mathbf{p}_2'$, $\mathbf{q}_2$ and $\mathbf{p}_2$ using the various delta functions yields

$$\frac{1}{2}\int_{-\infty}^{\infty} dt \langle [H_I^0(t), [H_I^0(0), \mathcal{D}_{ij}^0(\mathbf{k})]]\rangle$$
$$= \frac{1}{4}(2\pi)^3\delta^3(0) \int d\mathbf{q}\, d\mathbf{q}'\, d\mathbf{p}'(2\pi)^4\delta^4(q' + p' - q - k)\mathcal{M}(q'r', p's_1', qr, ks_1)\mathcal{M}(qr, ks_2', q'r', p's_2)$$
$$\times [n_e(\mathbf{q})\delta_{is_1}\delta_{s_2 s_1'}\rho_{s_2' j}(\mathbf{k}) - n_e(\mathbf{q}')\delta_{is_1}\delta_{js_2'}\rho_{s_1' s_2}(\mathbf{p}')]$$
$$+ \frac{1}{4}(2\pi)^3\delta^3(0) \int d\mathbf{q}\, d\mathbf{q}'\, d\mathbf{p}(2\pi)^4\delta^4(q' + k - q - p)\mathcal{M}(q'r', ks_1', qr, ps_1)\mathcal{M}(qr, ps_2', q'r', ks_2)$$
$$\times [n_e(\mathbf{q}')\delta_{js_1'}\delta_{s_1 s_2'}\rho_{is_2}(\mathbf{k}) - n_e(\mathbf{q})\delta_{js_1'}\delta_{is_2}\rho_{s_2' s_1}(\mathbf{p})]. \quad (4.17)$$

The subscript "1" on all momentum variables has been dropped for notational simplicity. All terms quadratic in the electron phase-space density have been dropped since for all cosmological scenarios this number is negligible compared to unity; all terms quadratic in the photon density matrix cancel exactly. By relabeling the integration variables and spin indices (implicitly summed over) in the second integral, Eq. (4.17) reduces to

$$\frac{1}{2}\int_{-\infty}^{\infty} dt \langle [H_I^0(t), [H_I^0(0), \mathcal{D}_{ij}^0(\mathbf{k})]]\rangle$$
$$= \frac{1}{4}(2\pi)^3\delta^3(0) \int d\mathbf{q}\, d\mathbf{q}'\, d\mathbf{p}'(2\pi)^4\delta^4(q' + p' - q - k)\mathcal{M}(q'r', p's_1', qr, ks_1)\mathcal{M}(qr, ks_2', q'r', p's_2)$$
$$\times \left[n_e(\mathbf{q})\delta_{s_2 s_1'}\left(\delta_{is_1}\rho_{s_2' j}(\mathbf{k}) + \delta_{js_2'}\rho_{is_1}(\mathbf{k})\right) - 2n_e(\mathbf{q}')\delta_{is_1}\delta_{js_2'}\rho_{s_1' s_2}(\mathbf{p}')\right]. \quad (4.18)$$

This equation is an essentially exact expression for the collision term in the case of the microwave background: the approximation that the duration of the Compton scattering be small compared to the time between scatterings is eminently satisfied for any cosmological scenario, and assuming the electrons to be unpolarized is essentially exact unless magnetic fields become important at some epoch.

Evaluating the matrix elements and performing the integrals in Eq. (4.18) is a straightforward process. This paper is concerned with the first-order perturbations away from a perfectly homogeneous and isotropic universe, and the scattering term will be explicitly calculated to first order. Evaluating the matrix elements involves standard techniques of quantum field theory and yields the familiar Compton cross-section to lowest order:

$$\sum_{rr'} \mathcal{M}(q'r', ps_1', qr, ks_1)\mathcal{M}(qr, ks_2', q'r', ps_2)$$
$$= 2e^4 \left[ \left( \frac{q \cdot k}{q \cdot p} + \frac{q \cdot p}{q \cdot k} \right) \left( \varepsilon_{s_1}(\mathbf{k}) \cdot \varepsilon_{s_1'}(\mathbf{p}) \varepsilon_{s_2}(\mathbf{p}) \cdot \varepsilon_{s_2'}(\mathbf{k}) - \varepsilon_{s_1}(\mathbf{k}) \cdot \varepsilon_{s_2}(\mathbf{p}) \varepsilon_{s_1'}(\mathbf{p}) \cdot \varepsilon_{s_2'}(\mathbf{k}) + \delta_{s_1 s_2'} \delta_{s_1' s_2} \right) \right.$$
$$\left. + 2 \left( \varepsilon_{s_1}(\mathbf{k}) \cdot \varepsilon_{s_1'}(\mathbf{p}) \varepsilon_{s_2}(\mathbf{p}) \cdot \varepsilon_{s_2'}(\mathbf{k}) + \varepsilon_{s_1}(\mathbf{k}) \cdot \varepsilon_{s_2}(\mathbf{p}) \varepsilon_{s_1'}(\mathbf{p}) \cdot \varepsilon_{s_2'}(\mathbf{k}) - \delta_{s_1 s_2'} \delta_{s_1' s_2} \right) \right]. \quad (4.19)$$

The following subsection then obtains the general Boltzmann equation for the photon density matrix to first order in terms of the photon energy and polarization vectors.

### D. Scattering Term to First Order

Now we proceed to evaluate Eq. (4.18) to lowest order in scattering kinematics. After substituting the matrix element Eq. (4.19) into Eq. (4.18), the Boltzmann equation (3.12), now explicitly including spatial dependence, becomes

$$\frac{d}{dt}\rho_{ij}(\mathbf{x},\mathbf{k}) = -\frac{e^4}{16m^2 k} \int d\mathbf{q}\, d\mathbf{p} \frac{m}{E(\mathbf{q}+\mathbf{k}-\mathbf{p})} (2\pi)\delta\left(E(\mathbf{q}+\mathbf{k}-\mathbf{p}) + p - E(\mathbf{q}) - k\right)$$
$$\times \left( n_e(\mathbf{x},\mathbf{q})\delta_{s_2 s_1'}\left(\delta_{is_1}\rho_{s_2' j}(\mathbf{x},\mathbf{k}) + \delta_{js_2'}\rho_{is_1}(\mathbf{x},\mathbf{k})\right) - 2n_e(\mathbf{x},\mathbf{q}')\delta_{is_1}\delta_{js_2'}\rho_{s_1' s_2}(\mathbf{x},\mathbf{p}')\right)$$
$$\times \left[ \left( \frac{q \cdot p}{q \cdot k} + \frac{q \cdot k}{q \cdot p} \right) \left( \varepsilon_{s_1'}(\mathbf{p}) \cdot \varepsilon_{s_1}(\mathbf{k}) \varepsilon_{s_2'}(\mathbf{k}) \cdot \varepsilon_{s_2}(\mathbf{p}) - \varepsilon_{s_1}(\mathbf{k}) \cdot \varepsilon_{s_2}(\mathbf{p}) \varepsilon_{s_1'}(\mathbf{p}) \cdot \varepsilon_{s_2'}(\mathbf{k}) + \delta_{s_1' s_2} \delta_{s_1 s_2'} \right) \right.$$
$$\left. + 2 \left( \varepsilon_{s_1'}(\mathbf{p}) \cdot \varepsilon_{s_1}(\mathbf{k}) \varepsilon_{s_2'}(\mathbf{k}) \cdot \varepsilon_{s_2}(\mathbf{p}) + \varepsilon_{s_1}(\mathbf{k}) \cdot \varepsilon_{s_2}(\mathbf{p}) \varepsilon_{s_1'}(\mathbf{p}) \cdot \varepsilon_{s_2'}(\mathbf{k}) - \delta_{s_1' s_2}\delta_{s_1 s_2'} \right) \right] \quad (4.20)$$

where $E(\mathbf{q}) = (\mathbf{q}^2 + m^2)^{1/2}$ is the energy of an electron with momentum $\mathbf{q}$. The electrons are described by an unpolarized thermal Maxwell-Boltzmann distribution:

$$n_e(\mathbf{x}, \mathbf{q}) = n_e(\mathbf{x}) \left( \frac{2\pi}{mT_e} \right)^{3/2} \exp\left[ -\frac{(\mathbf{q} - m\mathbf{v}(\mathbf{x}))^2}{2mT_e} \right], \quad (4.21)$$

with $T_e$ the electron temperature and $\mathbf{v}(\mathbf{x})$ the electron bulk velocity. Useful integrals of the electron distribution are

$$\int \frac{d^3\mathbf{q}}{(2\pi)^3} n_e(\mathbf{x},\mathbf{q}) = n_e(\mathbf{x}), \quad (4.22\text{a})$$

$$\int \frac{d^3\mathbf{q}}{(2\pi)^3} q_i n_e(\mathbf{x},\mathbf{q}) = m v_i(\mathbf{x}) n_e(\mathbf{x}). \quad (4.22\text{b})$$

For relevant cosmological situations, the kinetic energies of the electrons and photons are negligible compared to the electron mass, implying that the energy transfer in a Compton scattering event is small compared to the characteristic photon energy: $p \ll m$, $q \ll m$ (with obvious abbreviations $p = p^0 = |\mathbf{p}|$ and $q = |\mathbf{q}|$). Furthermore, if the electron and photon temperatures are comparable, $p \ll q$. We expand the various functions in Eq. (4.20) in terms of $p/q$ and $q/m$, using the following asymptotic expansions:

$$E(\mathbf{q}+\mathbf{Q}) \sim m \left[ 1 + \frac{\mathbf{q}^2}{m^2} + \frac{\mathbf{q} \cdot \mathbf{Q}}{m^2} + \frac{\mathbf{Q}^2}{2m^2} + \cdots \right], \quad (4.23\text{a})$$

$$n_e(\mathbf{q}+\mathbf{Q}) \sim n_e(\mathbf{q}) \left[ 1 - \frac{\mathbf{Q} \cdot (\mathbf{q} - m\mathbf{v})}{mT_e} - \frac{\mathbf{Q}^2}{2mT_e} + \cdots \right], \quad (4.23\text{b})$$

$$\delta\left(k - p + E(\mathbf{q}) - E(\mathbf{q} + \mathbf{k} - \mathbf{p})\right) \sim \delta(k-p) + \frac{(\mathbf{k} - \mathbf{p}) \cdot \mathbf{q}}{m} \frac{\partial \delta(k-p)}{\partial p} + \cdots, \tag{4.23c}$$

where in the last expression the derivative of the delta functional is defined through integration by parts [33]. Writing out the polarization sums explicitly yields the equation

$$\frac{d}{dt} \rho_{ij}(\mathbf{x}, \mathbf{k}) = \frac{e^4 n_e(\mathbf{x})}{16\pi m^2 k} \int_0^\infty dp\, p \int \frac{d\Omega}{4\pi} \left[ \delta(k-p) + (\mathbf{k} - \mathbf{p}) \cdot \mathbf{v}(\mathbf{x}) \frac{\partial \delta(k-p)}{\partial p} \right]$$

$$\times \left\{ -2 \left(\frac{p}{k} + \frac{k}{p}\right) \rho_{ij}(\mathbf{x}, \mathbf{k}) + 4\hat{\mathbf{p}} \cdot \hat{\varepsilon}_i(\mathbf{k})\, \hat{\mathbf{p}} \cdot \hat{\varepsilon}_1(\mathbf{k}) \rho_{1j}(\mathbf{x}, \mathbf{k}) + 4\hat{\mathbf{p}} \cdot \hat{\varepsilon}_i(\mathbf{k})\, \hat{\mathbf{p}} \cdot \hat{\varepsilon}_2(\mathbf{k}) \rho_{2j}(\mathbf{x}, \mathbf{k}) \right.$$

$$+ \left(\frac{p}{k} + \frac{k}{p} - 2\right) \delta_{ij} \left(\rho_{11}(\mathbf{x}, \mathbf{p}) + \rho_{22}(\mathbf{x}, \mathbf{p})\right)$$

$$+ \left(\frac{p}{k} + \frac{k}{p}\right) \left(\varepsilon_i(\mathbf{k}) \cdot \varepsilon_1(\mathbf{p})\, \varepsilon_j(\mathbf{k}) \cdot \varepsilon_2(\mathbf{p}) - \varepsilon_i(\mathbf{k}) \cdot \varepsilon_2(\mathbf{p})\, \varepsilon_j(\mathbf{k}) \cdot \varepsilon_1(\mathbf{p})\right) \left(\rho_{12}(\mathbf{x}, \mathbf{p}) - \rho_{21}(\mathbf{x}, \mathbf{p})\right)$$

$$+ 2\left(\varepsilon_i(\mathbf{k}) \cdot \varepsilon_1(\mathbf{p})\, \varepsilon_j(\mathbf{k}) \cdot \varepsilon_2(\mathbf{p}) + \varepsilon_i(\mathbf{k}) \cdot \varepsilon_2(\mathbf{p})\, \varepsilon_j(\mathbf{k}) \cdot \varepsilon_1(\mathbf{p})\right) \left(\rho_{12}(\mathbf{x}, \mathbf{p}) + \rho_{21}(\mathbf{x}, \mathbf{p})\right)$$

$$\left. + 4\varepsilon_i(\mathbf{k}) \cdot \varepsilon_1(\mathbf{p})\, \varepsilon_j(\mathbf{k}) \cdot \varepsilon_1(\mathbf{p}) \rho_{11}(\mathbf{x}, \mathbf{p}) + 4\varepsilon_i(\mathbf{k}) \cdot \varepsilon_2(\mathbf{p})\, \varepsilon_j(\mathbf{k}) \cdot \varepsilon_2(\mathbf{p}) \rho_{22}(\mathbf{x}, \mathbf{p}) \right\} \tag{4.24}$$

Here the photon momentum integral has been rewritten as an energy integral and an angular integral over the momentum direction. This is the basic equation describing the evolution of the photon density matrix to first order in the kinematic variables. By rewriting the momentum and polarization vectors in a spherical coordinate basis and incorporating the velocity-dependent term into the left-hand side, the equation becomes equivalent to Chandrasekhar's radiative transfer formalism; *cf.* Chapter 1, Eq. (212) of Ref. [21]. Before performing the final angular integrals, we must consider the left side of the equation for the particular space-time geometry in which we are interested, which determines the azimuthal dependence of $\rho$. The left side of the equation will be analyzed in the following section, and then in Section VI we perform the remaining momentum integrals to complete the evaluation of the right side.

## V. THE GENERAL-RELATIVISTIC LIOUVILLE EQUATION

The last section has analyzed the right side of the Boltzmann equation, Eq. (3.12); we now turn to the left side, describing the propogation of photons in the background space-time. The Boltzmann equation with no collision term on the right side is the Liouville equation, describing the evolution of a collisionless system's phase space distribution. Writing the equation has already assumed definition of a set of space-like hypersurfaces; that is, the equation contains an explicit time derivative. The background space-time here will be the canonical Friedmann-Robertson-Walker (FRW) space-time. In this paper only the flat case will be considered; techniques pertaining to open universes have also been extensively developed [34]. Scalar and tensor metric perturbations are added to the flat background space-time; we neglect vector perturbations, which kinematically decay and are unimportant unless a continual source of vector perturbations, such as topological defects, exists. The metric we consider is

$$g_{00} = 1 + 2\Phi(\mathbf{x}, t), \qquad g_{0i} = 0, \tag{5.1}$$

$$g_{ij} = -a^2(t)\left[(1 - 2\Psi(\mathbf{x}, t))\delta_{ij} + h_{ij}(\mathbf{x}, t)\right]. \tag{5.2}$$

In this section, Greek subscripts refer to space-time indices running from 0 to 3 while Roman subscripts refer to spatial indices running from 1 to 3. The function $a(t)$ is the usual cosmological scale factor. The scalar perturbations, defined in the longitudinal gauge, are given by the two scalar function $\Phi$ and $\Psi$ [26]. The metric perturbations $h_{ij}$ are defined in the transverse-traceless gauge and are subject to the constraints

$$h_i{}^i = 0, \qquad \partial^j h_{ij} = 0. \tag{5.3}$$

With the perturbations defined in this way, no residual gauge freedom remains, in contrast to the more conventional synchronous gauge condition. A second advantage of the present definition is that in the Newtonian limit, the metric perturbation $\Phi$ simply corresponds to the Newtonian potential. The inverse metric to first order in the perturbations is

$$g^{00} = 1 - 2\Phi(\mathbf{x}, t), \qquad g^{0i} = 0, \tag{5.4}$$

$$g^{ij} = -\frac{1}{a^2(t)}\left[(1 + 2\Psi(\mathbf{x}, t))\delta^{ij} - h^{ij}(\mathbf{x}, t)\right]. \tag{5.5}$$

In this section we will consider the Liouville equation to first order in the metric perturbations; second-order treatments have been undertaken in Ref. [35].

Photons are described by space-time coordinate $x^\mu$ and four-momentum $k^\mu$. Our coordinate system has $x^0 \equiv t$; the photon momentum satisfies

$$\frac{k^i}{k^0} = \frac{dx^i}{dt}. \tag{5.6}$$

The photons obey the geodesic equation

$$\frac{d^2 x^\mu}{d\lambda^2} + \Gamma^\mu{}_{\alpha\beta}\frac{dx^\alpha}{d\lambda}\frac{dx^\beta}{d\lambda} = 0, \tag{5.7}$$

$$g_{\mu\nu}\frac{dx^\mu}{d\lambda}\frac{dx^\nu}{d\lambda} = 0, \tag{5.8}$$

where $\lambda$ is an affine parameter along the photon geodesic, which may be defined so that $dt/d\lambda \equiv dx^0/d\lambda = k^0$; thus $dx^i/d\lambda = k^i$ using Eq. (5.6). Therefore, using the definition of the Christoffel symbol $\Gamma$,

$$\frac{dk^\mu}{dt} = g^{\mu\nu}\left(\frac{1}{2}\frac{\partial g_{\alpha\beta}}{\partial x^\nu} - \frac{\partial g_{\nu\alpha}}{\partial x^\beta}\right)\frac{k^\alpha k^\beta}{k^0} \tag{5.9}$$

with the geodesic condition $k^\mu k_\mu = 0$.

Liouville's equation for any phase-space distribution function $f$ is

$$\frac{df}{dt} = \frac{\partial f}{\partial x^\mu}\frac{dx^\mu}{dt} + \frac{\partial f}{\partial k^\mu}\frac{dk^\mu}{dt} = \frac{\partial f}{\partial t} + \frac{\partial f}{\partial x^i}\frac{k^i}{k^0} + \frac{\partial f}{\partial k^\mu}\frac{dk^\mu}{dt} = 0. \tag{5.10}$$

We change to a convenient choice of momentum variables, the photon energy in a local orthonormal frame $k \equiv (-k^i k_i)^{1/2}$ and the unit vector $\hat{k}^i$. A local observer at rest with respect to the cosmic coordinate system will measure the photon to have energy $k$; this is the energy appearing in the collision term on the right side of the Boltzmann Equation. The wave vector is thus given by

$$k^i = \frac{1}{a}k\hat{k}^i\left(1 + \Psi - \frac{1}{2}\hat{k}^m\hat{k}^n h_{mn}\right), \tag{5.11}$$

and the Liouville equation in the new variables is

$$\frac{df}{dt} = \frac{\partial f}{\partial t} + \frac{\partial f}{\partial x^i}\frac{k^i}{k^0} + \frac{\partial f}{\partial k}\frac{dk}{dt} + \frac{\partial f}{\partial \hat{k}^i}\frac{d\hat{k}^i}{dt} = 0. \tag{5.12}$$

Now it is a straightforward matter to calculate the derivatives $dk/dt$ and $d\hat{k}^i/dt$. Substituting $k^0 = k(1 - \Phi)$ and Eq. (5.11) into the zero component of Eq. (5.9) gives to first order

$$\frac{dk}{dt} = -k\left[\frac{\dot{a}}{a} - \frac{\partial \Psi}{\partial t} + \frac{\hat{k}^i}{a}\frac{\partial \Phi}{\partial x^i} + \frac{1}{2}\hat{k}^i\hat{k}^j\frac{\partial h_{ij}}{\partial t}\right]. \tag{5.13}$$

The derivative of $\hat{k}^i$ is most easily computed by differentiating Eq. (5.11) and equating this result with Eq. (5.9); it is straightforward to show that $d\hat{k}^i/dt$ has no lowest-order terms. Physically, this is because geodesics are straight lines in the unperturbed metric. Since $\partial f/\partial \hat{k}^i$ is itself linear in the metric perturbations, the final term of the Liouville equation drops out to first order. Expanding the distribution function as

$$f(\mathbf{x}, k, \hat{k}, t) = f^{(0)}(k, t) + f^{(1)}(\mathbf{x}, k, \hat{k}, t) \tag{5.14}$$

leads to the zeroth-order equation

$$\frac{\partial f^{(0)}}{\partial t} - \frac{\dot{a}}{a} k \frac{\partial f^{(0)}}{\partial k} = 0 \qquad (5.15)$$

whose solution is simply $f^{(0)}(k,t) = f^{(0)}(ka)$, which is just the uniform redshift of the spectrum with cosmic expansion. The first-order Liouville equation is

$$\frac{\partial f^{(1)}}{\partial t} + \frac{\partial f^{(1)}}{\partial x^i} \frac{\hat{k}^i}{a} - \frac{\dot{a}}{a} k \frac{\partial f^{(1)}}{\partial k} + \frac{\partial f^{(0)}}{\partial k} k \left[ \frac{\partial \Psi}{\partial t} - \frac{\hat{k}^i}{a} \frac{\partial \Phi}{\partial x^i} - \frac{1}{2} \hat{k}^i \hat{k}^j \frac{\partial h_{ij}}{\partial t} \right] = 0. \qquad (5.16)$$

Note that $k$ in this equation is the physical, not the comoving, photon wave number.

The terms in the Liouville equation depending directly on the metric perturbations determine the form of the directional dependence of the distribution function. A Fourier transform over the spatial dependence of the equation gives for the Boltzmann equation

$$\frac{\partial}{\partial t} f^{(1)}(\mathbf{K}, k, \hat{\mathbf{k}}) + \frac{i}{a} (\mathbf{K} \cdot \hat{\mathbf{k}}) f^{(1)}(\mathbf{K}, k, \hat{\mathbf{k}}) - \frac{\dot{a}}{a} k \frac{\partial}{\partial k} f^{(1)}(\mathbf{K}, k, \hat{\mathbf{k}})$$
$$- \frac{\partial f^{(0)}(k)}{\partial k} k \left[ \frac{\partial}{\partial t} \Phi(\mathbf{K}) - \frac{\partial}{\partial t} \Psi(\mathbf{K}) + \frac{i}{a} (\mathbf{K} \cdot \hat{\mathbf{k}}) \Phi(\mathbf{K}) + \frac{1}{2} \hat{k}^i \hat{k}^j \frac{\partial}{\partial t} h_{ij}(\mathbf{K}) \right] = C(\mathbf{K}, k, \hat{\mathbf{k}}) \qquad (5.17)$$

where $C$ represents the collision term on the right side. For scalar perturbations, the previous section shows the right side contains source terms proportional to $\hat{\mathbf{k}} \cdot \mathbf{v}$, where $v$ is the local velocity of the electrons. But for scalar perturbations, $\mathbf{v} \propto \mathbf{K}$ [26], so if we choose spherical coordinates for $\hat{\mathbf{k}}$ with axis in the $\mathbf{K}$ direction, then $f^{(1)}$ is manifestly independent of the azimuthal angle $\phi$; in other words,

$$f^{(1)}(\mathbf{K}, k, \hat{\mathbf{k}}) = f^{(1)}(\mathbf{K}, k, \theta) \qquad (5.18)$$

for scalar perturbations.

Tensor perturbations do depend on $\phi$, but in a simple manner. We neglect any electron velocity $\mathbf{v}$ arising from tensor perturbations as corrections to the scalar-induced velocity. The $\phi$ dependence of the distribution function is determined by the perturbation term, which can be written as

$$\hat{k}^i \hat{k}^j \frac{\partial}{\partial t} h_{ij}(\mathbf{K}, t) = \hat{k}^i \hat{k}^j \left( \dot{h}^+(\mathbf{K}, t) e_{ij}^+(\mathbf{K}) + \dot{h}^\times(\mathbf{K}, t) e_{ij}^\times(\mathbf{K}) \right), \qquad (5.19)$$

where $e_{ij}^+$ and $e_{ij}^\times$ are polarization tensors for the plus and cross polarizations of the gravity wave. Again, choose spherical coordinates with the $z$-axis pointing in the direction of $\mathbf{K}$. In this coordinate system the polarization tensors are given by $e_{xx}^+ = -e_{yy}^+ = 1$ and $e_{xy}^\times = e_{yx}^\times = 1$, with the other components zero. Contraction of the unit vectors with the polarization tensors gives

$$\hat{k}^i \hat{k}^j e_{ij}^+ = \sin^2 \theta \cos^2 \phi - \sin^2 \theta \sin^2 \phi = \sin^2 \theta \cos 2\phi,$$
$$\hat{k}^i \hat{k}^j e_{ij}^\times = 2 \sin^2 \theta \cos \phi \sin \phi = \sin^2 \theta \sin 2\phi. \qquad (5.20)$$

Therefore, for a given plane wave component of a metric tensor perturbation,

$$f^{(1)}(\mathbf{K}, k, \hat{\mathbf{k}}) = f^{(1)}(\mathbf{K}, k, \theta) \cos 2\phi \qquad (5.21\text{a})$$

for the plus polarization of the gravity wave and

$$f^{(1)}(\mathbf{K}, k, \hat{\mathbf{k}}) = f^{(1)}(\mathbf{K}, k, \theta) \sin 2\phi \qquad (5.21\text{b})$$

for the cross polarization.

## VI. COMPLETE POLARIZATION EQUATIONS

Now we have assembled all the ingredients for deriving the final polarization evolution equations: Eq. (4.24) with the perturbation expansion for $\rho_{ij}$ given by Eq. (5.14) and the angular dependence of $\rho_{ij}$ given by Eqs. (5.18) and (5.21). Note the unperturbed photon density matrix satisfies $\rho_{11}^{(0)} = \rho_{22}^{(0)}$ and $\rho_{12}^{(0)} = \rho_{21}^{(0)} = 0$, since it represents a uniform unpolarized blackbody spectrum. The lowest-order term on the right side of Eq. (4.24) is zero; the resulting equation for $\rho^{(0)}$ gives the uniform shift of the photon spectrum with scale factor in an expanding universe, Eq. (5.15). The first order term gives

$$\begin{aligned}
\frac{d}{dt}\rho_{ij}^{(1)}(\mathbf{K},\mathbf{k}) = \frac{e^4 \bar{n}_e}{16\pi m^2 k} \int_0^\infty dp\, p \int \frac{d\Omega}{4\pi} &\bigg\{ \delta(k-p)\bigg[ -2\left(\frac{p}{k}+\frac{k}{p}\right)\rho_{ij}^{(1)}(\mathbf{K},\mathbf{k}) \\
&+ 4\hat{\mathbf{p}}\cdot\hat{\varepsilon}_i(\mathbf{k})\,\hat{\mathbf{p}}\cdot\hat{\varepsilon}_1(\mathbf{k})\rho_{1j}^{(1)}(\mathbf{K},\mathbf{k}) + 4\hat{\mathbf{p}}\cdot\hat{\varepsilon}_i(\mathbf{k})\,\hat{\mathbf{p}}\cdot\hat{\varepsilon}_2(\mathbf{k})\rho_{2j}^{(1)}(\mathbf{K},\mathbf{k}) \\
&+ \left(\frac{p}{k}+\frac{k}{p}\right)\Big(\varepsilon_i(\mathbf{k})\cdot\varepsilon_1(\mathbf{p})\,\varepsilon_j(\mathbf{k})\cdot\varepsilon_2(\mathbf{p}) - \varepsilon_i(\mathbf{k})\cdot\varepsilon_2(\mathbf{p})\,\varepsilon_j(\mathbf{k})\cdot\varepsilon_1(\mathbf{p})\Big)\Big(\rho_{12}^{(1)}(\mathbf{K},\mathbf{p}) - \rho_{21}^{(1)}(\mathbf{K},\mathbf{p})\Big) \\
&+ 2\Big(\varepsilon_i(\mathbf{k})\cdot\varepsilon_1(\mathbf{p})\,\varepsilon_j(\mathbf{k})\cdot\varepsilon_2(\mathbf{p}) + \varepsilon_i(\mathbf{k})\cdot\varepsilon_2(\mathbf{p})\,\varepsilon_j(\mathbf{k})\cdot\varepsilon_1(\mathbf{p})\Big)\Big(\rho_{12}^{(1)}(\mathbf{K},\mathbf{p}) + \rho_{21}^{(1)}(\mathbf{K},\mathbf{p})\Big) \\
&+ 4\varepsilon_i(\mathbf{k})\cdot\varepsilon_1(\mathbf{p})\,\varepsilon_j(\mathbf{k})\cdot\varepsilon_1(\mathbf{p})\rho_{11}^{(1)}(\mathbf{K},\mathbf{p}) + 4\varepsilon_i(\mathbf{k})\cdot\varepsilon_2(\mathbf{p})\,\varepsilon_j(\mathbf{k})\cdot\varepsilon_2(\mathbf{p})\rho_{22}^{(1)}(\mathbf{K},\mathbf{p})\bigg] \\
&+ (\mathbf{k}-\mathbf{p})\cdot\mathbf{v}(\mathbf{K})\frac{\partial\delta(k-p)}{\partial p}\bigg[ -2\left(\frac{p}{k}+\frac{k}{p}\right)\rho_{ij}^{(0)}(k) + \left(\frac{p}{k}+\frac{k}{p}-2\right)\delta_{ij}\Big(\rho_{11}^{(0)}(p)+\rho_{22}^{(0)}(p)\Big) \\
&+ 4\hat{\mathbf{p}}\cdot\hat{\varepsilon}_i(\mathbf{k})\,\hat{\mathbf{p}}\cdot\hat{\varepsilon}_1(\mathbf{k})\rho_{1j}^{(0)}(k) + 4\hat{\mathbf{p}}\cdot\hat{\varepsilon}_i(\mathbf{k})\,\hat{\mathbf{p}}\cdot\hat{\varepsilon}_2(\mathbf{k})\rho_{2j}^{(0)}(k) \\
&+ 4\varepsilon_i(\mathbf{k})\cdot\varepsilon_1(\mathbf{p})\,\varepsilon_j(\mathbf{k})\cdot\varepsilon_1(\mathbf{p})\rho_{11}^{(0)}(p) + 4\varepsilon_i(\mathbf{k})\cdot\varepsilon_2(\mathbf{p})\,\varepsilon_j(\mathbf{k})\cdot\varepsilon_2(\mathbf{p})\rho_{22}^{(0)}(p)\bigg]\bigg\}.
\end{aligned} \quad (6.1)$$

As in the previous section, $\mathbf{K}$ is the Fourier conjugate of $\mathbf{x}$, and $\bar{n}_e$ is the mean electron density which is constant to lowest order. The remainder of this section evaluates the remaining angular integrals in this expression and converts the equations for the density matrix elements to equations for the brightness of each Stokes parameter.

To evaluate the angular integrals most conveniently, choose the $z$-axis of the spherical coordinate system to coincide with $\mathbf{K}$, independently for each Fourier mode. Note that on transforming back to real-space coordinates, care must be taken because the density matrix is not invariant under a change of basis (see Sec. II). The basis for the photon direction and polarization vectors is taken to be

$$\begin{aligned}
\hat{k}_x &= \sin\theta\cos\phi & \hat{\varepsilon}_{1x}(\mathbf{k}) &= \cos\theta\cos\phi & \hat{\varepsilon}_{2x}(\mathbf{k}) &= -\sin\phi \\
\hat{k}_y &= \sin\theta\sin\phi & \hat{\varepsilon}_{1y}(\mathbf{k}) &= \cos\theta\sin\phi & \hat{\varepsilon}_{2y}(\mathbf{k}) &= \cos\phi \\
\hat{k}_z &= \cos\theta & \hat{\varepsilon}_{1z}(\mathbf{k}) &= -\sin\theta & \hat{\varepsilon}_{2z}(\mathbf{k}) &= 0.
\end{aligned} \quad (6.2)$$

The same definition is used for $\mathbf{p}$ and its associated polarization vectors, with $\theta \to \theta'$ and $\phi \to \phi'$. The angular integral in Eq. (6.1) is over $\theta'$ and $\phi'$, and the various dot products are given by

$$\begin{aligned}
\hat{\mathbf{p}}'\cdot\hat{\varepsilon}_1(\mathbf{k}) &= \sin\theta'\cos\theta\cos(\phi'-\phi) - \cos\theta'\sin\theta, \\
\hat{\mathbf{p}}'\cdot\hat{\varepsilon}_2(\mathbf{k}) &= \sin\theta'\sin(\phi'-\phi), \\
\hat{\varepsilon}_1(\mathbf{k})\cdot\hat{\varepsilon}_1(\mathbf{p}) &= \cos\theta\cos\theta'\cos(\phi'-\phi) + \sin\theta\sin\theta', \\
\hat{\varepsilon}_1(\mathbf{k})\cdot\hat{\varepsilon}_2(\mathbf{p}) &= -\cos\theta\sin(\phi'-\phi), \\
\hat{\varepsilon}_2(\mathbf{k})\cdot\hat{\varepsilon}_1(\mathbf{p}) &= \cos\theta'\sin(\phi'-\phi), \\
\hat{\varepsilon}_2(\mathbf{k})\cdot\hat{\varepsilon}_2(\mathbf{p}) &= \cos(\phi'-\phi).
\end{aligned} \quad (6.3)$$

Two additional convenient abbreviations are $\mu' \equiv \hat{\mathbf{v}}\cdot\hat{\mathbf{p}} = \cos\theta'$ and $\mu \equiv \hat{\mathbf{v}}\cdot\hat{\mathbf{k}} = \cos\theta$. Now the angular integrals are straightforward, resulting in expressions like

$$\begin{aligned}
\frac{d}{dt}\rho_{11}^{(1)}(k,\mu) = -\frac{e^4\bar{n}_e}{16\pi m^2}\bigg[&\frac{8}{3}\rho_{11}^{(1)}(k,\mu) - \left(4\mu^2 - \frac{8}{3}\right)\int_{-1}^1 \frac{d\mu'}{2} P_2(\mu')\rho_{11}^{(1)}(k,\mu') - \left(\frac{8}{3} - 2\mu^2\right)\int_{-1}^1 \frac{d\mu'}{2}\rho_{11}^{(1)}(k,\mu') \\
&- 2\mu^2 \int_{-1}^1 \frac{d\mu'}{2}\rho_{22}^{(1)}(k,\mu') + kv\mu\left(\frac{8}{3}-2\mu^2\right)\frac{\partial\rho_{11}^{(0)}}{\partial k} + 2kv\mu^3 \frac{\partial\rho_{22}^{(0)}}{\partial k}\bigg]
\end{aligned} \quad (6.4)$$

for scalar perturbations, where dependence on the Fourier mode $\mathbf{K}$ is implicit. In solving the evolution equations, it is convenient to split the density-matrix perturbation into two parts, one due to the scalar metric perturbations and one due to the tensor metric perturbation. In making this split, the bulk velocity **v** is entirely attributed to the scalar perturbations.

For the final set of evolution equations, we change variables to comoving wave number $q = ka$, and convert the density matrix elements to Stokes parameter brightness perturbations:

$$\Delta_I^i(\mathbf{K}, q, \mu) \equiv \left[\frac{q}{4}\frac{\partial \rho_{11}^{(0)}(q)}{\partial q}\right]^{-1} \left(\rho_{11}^{(1)}(\mathbf{K}, q, \mu) + \rho_{22}^{(1)}(\mathbf{K}, q, \mu)\right), \tag{6.5a}$$

$$\Delta_Q^i(\mathbf{K}, q, \mu) \equiv \left[\frac{q}{4}\frac{\partial \rho_{11}^{(0)}(q)}{\partial q}\right]^{-1} \left(\rho_{11}^{(1)}(\mathbf{K}, q, \mu) - \rho_{22}^{(1)}(\mathbf{K}, q, \mu)\right), \tag{6.5b}$$

$$\Delta_U^i(\mathbf{K}, q, \mu) \equiv \left[\frac{q}{4}\frac{\partial \rho_{11}^{(0)}(q)}{\partial q}\right]^{-1} \left(\rho_{12}^{(1)}(\mathbf{K}, q, \mu) + \rho_{21}^{(1)}(\mathbf{K}, q, \mu)\right), \tag{6.5c}$$

$$\Delta_V^i(\mathbf{K}, q, \mu) \equiv -i \left[\frac{q}{4}\frac{\partial \rho_{11}^{(0)}(q)}{\partial q}\right]^{-1} \left(\rho_{12}^{(1)}(\mathbf{K}, q, \mu) - \rho_{21}^{(1)}(\mathbf{K}, q, \mu)\right), \tag{6.5d}$$

where the superscript $i$ stands for $s, +$, or $\times$, representing the three types of possible metric perturbations: scalar and two polarizations of tensor. For linear perturbations considered here, the distribution function undergoes no spectral distortions and the perturbations are blackbody; in this case, $\Delta_I/4$ is just the temperature fluctuation $\delta T/T_0$ with $T_0$ the mean temperature. We also define moments of these variables:

$$\Delta_{Il}^i(q) \equiv \int_{-1}^{1} \frac{d\mu'}{2} P_l(\mu') \Delta_I^i(q, \mu') \tag{6.6}$$

where $P_l$ is the Legendre polynomial of order $l$. Note these moments are sometimes defined differently [36,37] as $\Delta(\mu) = \sum (-i)^l \Delta_l P_l$.

For scalar perturbations, the brightness is governed by the set of equations [17]

$$\frac{\partial \Delta_I^s}{\partial t} + \frac{1}{a}iK\mu\Delta_I^s + 4\left[\frac{\partial \Psi}{\partial t} - \frac{1}{a}iK\mu\Phi\right] = -\sigma_T \bar{n}_e \left[\Delta_I^s - \Delta_{I0}^s + 4v\mu - \frac{1}{2}P_2(\mu)(\Delta_{I2}^s + \Delta_{Q2}^s - \Delta_{Q0}^s)\right], \tag{6.7a}$$

$$\frac{\partial \Delta_Q^s}{\partial t} + \frac{1}{a}iK\mu\Delta_Q^s = -\sigma_T \bar{n}_e \left[\Delta_Q^s + \frac{1}{2}(1 - P_2(\mu))(\Delta_{I2}^s + \Delta_{Q2}^s - \Delta_{Q0}^s)\right], \tag{6.7b}$$

$$\frac{\partial \Delta_U^s}{\partial t} + \frac{1}{a}iK\mu\Delta_U^s = -\sigma_T \bar{n}_e \Delta_U^s, \tag{6.7c}$$

$$\frac{\partial \Delta_V^s}{\partial t} + \frac{1}{a}iK\mu\Delta_V^s = -\sigma_T \bar{n}_e \left[\Delta_V^s - \frac{3}{2}\mu\Delta_{V1}^s\right]. \tag{6.7d}$$

The evolution of the brightness thus does not depend on the direction of **K**, only on its magnitude; the brightness depends on the direction of **K** only through initial conditions, which factor out of the linear evolution equations. The equations for $U$ and $V$ have no source terms, so for each **K** the evolution leaves $U = 0$ and $V = 0$. The coordinate dependence of Eq. (2.3) gives a non-zero $U$ on transforming back to **x** space, but $V$ remains zero.

For tensor perturbations, the evolution equations take their simplest form after the coordinate transformation [38]

$$\Delta_I^+ = (1-\mu^2)\cos(2\phi)\tilde{\Delta}_I^+, \qquad \Delta_I^\times = (1-\mu^2)\sin(2\phi)\tilde{\Delta}_I^\times, \tag{6.8a}$$

$$\Delta_Q^+ = (1+\mu^2)\cos(2\phi)\tilde{\Delta}_Q^+, \qquad \Delta_Q^\times = (1+\mu^2)\sin(2\phi)\tilde{\Delta}_Q^\times, \tag{6.8b}$$

$$\Delta_U^+ = -2\mu\sin(2\phi)\tilde{\Delta}_U^+, \qquad \Delta_U^\times = 2\mu\cos(2\phi)\tilde{\Delta}_U^\times. \tag{6.8c}$$

The $\phi$ dependence is determined by Eqs. (5.21a) and (5.21b), and the $\mu$ dependence is chosen to simplify the final equations. After this change of variables, the brightness equations become [40]

$$\frac{\partial \tilde{\Delta}_I^+}{\partial t} + \frac{1}{a}iK\mu\tilde{\Delta}_I^+ - 2\frac{\partial h^+}{\partial t} = -\sigma_T \bar{n}_e \left(\tilde{\Delta}_I^+ + \tilde{\Lambda}^+\right), \tag{6.9a}$$

$$\frac{\partial \tilde{\Delta}_Q^+}{\partial t} + \frac{1}{a}iK\mu\tilde{\Delta}_Q^+ = -\sigma_T \bar{n}_e \left(\tilde{\Delta}_Q^+ - \tilde{\Lambda}^+\right), \tag{6.9b}$$

$$\tilde{\Delta}_U^+ = \tilde{\Delta}_Q^+, \tag{6.9c}$$

$$\frac{\partial \tilde{\Delta}_V^+}{\partial t} + \frac{1}{a} iK\mu \tilde{\Delta}_V^+ = -\sigma_T \bar{n}_e \tilde{\Delta}_V^+, \tag{6.9d}$$

$$\tilde{\Lambda}^+ \equiv -\frac{3}{70}\tilde{\Delta}_{I4}^+ + \frac{1}{7}\tilde{\Delta}_{I2}^+ - \frac{1}{10}\tilde{\Delta}_{I0}^+ + \frac{3}{70}\tilde{\Delta}_{Q4}^+ + \frac{6}{7}\tilde{\Delta}_{Q2}^+ + \frac{3}{5}\tilde{\Delta}_{Q0}^+ \tag{6.10}$$

The definition of $h^+$ is given in Eq. (5.19). The $\times$ tensor perturbation gives same equations. Again, $V = 0$ since it has no source term.

For a given cosmological scenario, which determines the metric perturbations, Eqs. (6.7) and (6.9) must be evolved numerically. This can be done efficiently by expanding the $\mu$ dependence of the brightness functions in terms of Legendre polynomials, Eq. (6.6), giving a large set of coupled ordinary differential equations [17]. Several detailed codes to calculate temperature fluctuations have been implemented using this scheme [4–7].

## VII. POWER SPECTRA

Numerical solution of the above transport equations gives the Fourier space brightness functions $\Delta_{Il}^s(K)$ and $\Delta_{Ql}^s(K)$ for scalar perturbations and $\tilde{\Delta}_{Il}^\epsilon(K)$ and $\tilde{\Delta}_{Ul}^\epsilon(K)$ for tensor perturbations, where $\epsilon$ represents the two gravity wave polarization states $+$ and $\times$. The temperature fluctuations in real space are then

$$\frac{T(\mathbf{x},\theta,\phi)}{T_0} = 1 + \frac{1}{4} \sum_{\mathbf{K}} \sum_l (2l+1) P_l(\cos\theta') e^{i\mathbf{K}\cdot\mathbf{x}}$$
$$\times \left[ \Delta_{Il}^s(\mathbf{K}) + \sin^2\theta' \cos 2\phi' \tilde{\Delta}_{Il}^+(\mathbf{K}) + \sin^2\theta' \sin 2\phi' \tilde{\Delta}_{Il}^\times(\mathbf{K}) \right], \tag{7.1}$$

where $T_0$ is the mean temperature of the microwave background, and $(\theta', \phi')$ represents the same direction as $(\theta, \phi)$ except in the coordinate system defined by the $\mathbf{K}$ direction.

The polarization is more complicated, because for each $\mathbf{K}$ mode the coordinate system in the direction $(\theta, \phi)$ has a different orientation; when the $Q$ and $U$ brightnesses are summed up, the axes must be rotated to the orientation in the $\mathbf{x}$ coordinate system, using Eq. (2.3). To determine this rotation angle for each $\mathbf{K}$ mode, refer to Fig. 3; the needed angle is labelled $\xi'$, the angle between the vectors $\hat{\theta}$ and $\hat{\theta}'$. Let the direction of $\mathbf{K}$ be denoted by $(\theta_K, \phi_K)$. On the unit sphere, the lengths of the sides of the spherical triangle $ABC$ are just the angles they subtend. The angle $\theta'$ is given by the law of cosines,

$$\cos\theta' = \cos\theta\cos\theta_K + \sin\theta\sin\theta_K \cos(\phi_K - \phi), \tag{7.2}$$

and the rotation angle is

$$\sin\xi' = \sin\theta_K \sin(\phi_K - \phi) \csc\theta'. \tag{7.3}$$

Then the $Q$ and $U$ brightnesses are given by

$$\frac{Q(\mathbf{x},\theta,\phi)}{T_0} = \frac{1}{4} \sum_{\mathbf{K}} \sum_l (2l+1) P_l(\cos\theta') e^{i\mathbf{K}\cdot\mathbf{x}}$$
$$\times \left[ \left( \Delta_{Ql}^s(\mathbf{K}) + (1+\cos^2\theta')\cos 2\phi' \tilde{\Delta}_{Ql}^+(\mathbf{K}) + (1+\cos^2\theta')\sin 2\phi' \tilde{\Delta}_{Ql}^\times(\mathbf{K}) \right) \cos 2\xi' \right.$$
$$\left. + \left( -2\cos\theta' \sin 2\phi' \tilde{\Delta}_{Ql}^+(\mathbf{K}) + 2\cos\theta' \cos 2\phi' \tilde{\Delta}_{Ql}^\times(\mathbf{K}) \right) \sin 2\xi' \right], \tag{7.4}$$

and $U$ the same except for the replacements $\cos\xi' \to -\sin\xi'$ and $\sin\xi' \to \cos\xi'$. From these two quantities, the polarization vector $\mathbf{P}$ follows from Eq. (2.4) as

$$\frac{\mathbf{P}(\mathbf{x},\theta,\phi)}{T_0} = \frac{1}{4\sqrt{2}} \left[ \hat{\theta}\sqrt{\Delta_P(\Delta_P + \Delta_Q)} + \hat{\phi}\frac{\Delta_U}{|\Delta_U|}\sqrt{\Delta_P(\Delta_P - \Delta_Q)} \right], \tag{7.5}$$

$$\Delta_P \equiv \sqrt{\Delta_Q^2 + \Delta_U^2}. \tag{7.6}$$

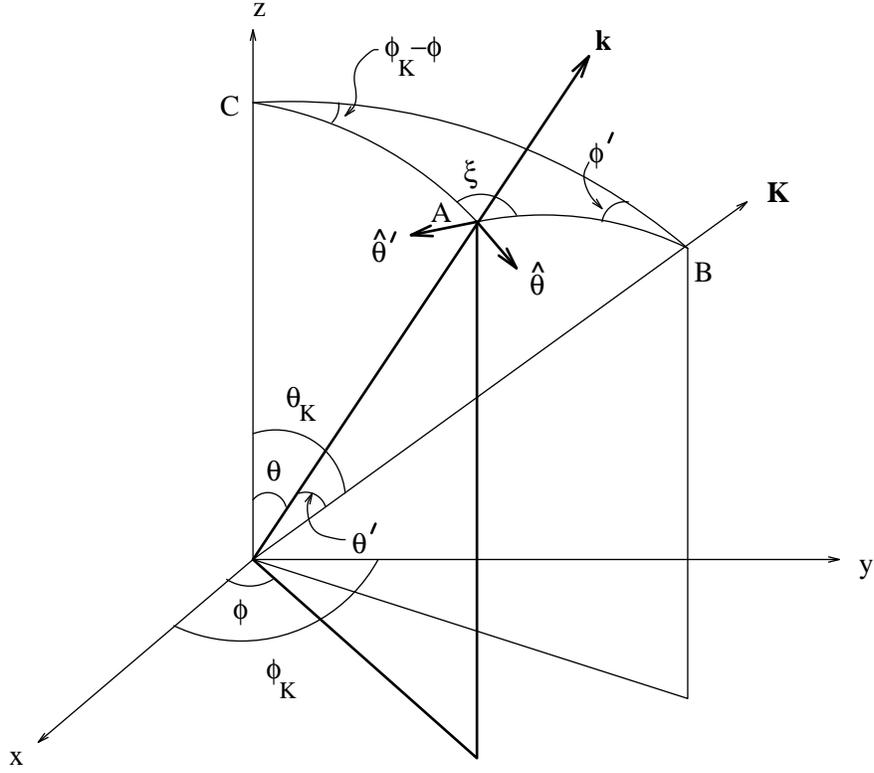

FIG. 3. Angles and directions for determining the orientation of different spherical coordinate bases at a given point.

The predictions of a given cosmological scenario are only statistical. The traditional statistical measure of temperature fluctuations is the angular power spectrum $C(\theta)$, defined by

$$C^{TT}(\theta) \equiv 1 + \left\langle \frac{T(\hat{\mathbf{q}}_1)}{T_0} \frac{T(\hat{\mathbf{q}}_2)}{T_0} \right\rangle, \qquad \hat{\mathbf{q}}_1 \cdot \hat{\mathbf{q}}_2 = \cos\theta, \tag{7.7}$$

where the angle brackets represent an ensemble average over initial conditions; this average can be replaced in calculations with an average over space, assuming ergodicity. Many analogous correlation functions for various polarization variables are possible:

$$C^{TP}(\theta) \equiv \left\langle \frac{T(\hat{\mathbf{q}}_1)}{T_0} \frac{P(\hat{\mathbf{q}}_2)}{T_0} \right\rangle, \tag{7.8a}$$

$$C^{PP}(\theta) \equiv \left\langle \frac{P(\hat{\mathbf{q}}_1)}{T_0} \frac{P(\hat{\mathbf{q}}_2)}{T_0} \right\rangle, \tag{7.8b}$$

$$C^{\mathcal{PP}}(\theta) \equiv \left\langle \frac{\mathbf{P}(\hat{\mathbf{q}}_1)}{T_0} \cdot \frac{\mathbf{P}(\hat{\mathbf{q}}_2)}{T_0} \right\rangle, \tag{7.8c}$$

$$C^{\mathcal{TP}}(\theta) \equiv \left\langle \frac{\mathbf{P}(\hat{\mathbf{q}}_1)}{T_0} \cdot \frac{\nabla T(\hat{\mathbf{q}}_2)}{T_0} \right\rangle, \tag{7.8d}$$

roughly in increasing order of difficulty to measure. The various combinations of Stokes parameters can also be used to form correlations:

$$C^{QQ}(\theta) \equiv \left\langle \frac{Q(\hat{\mathbf{q}}_1)}{T_0} \frac{Q(\hat{\mathbf{q}}_2)}{T_0} \right\rangle, \tag{7.9a}$$

$$C^{QT}(\theta) \equiv \left\langle \frac{Q(\hat{\mathbf{q}}_1)}{T_0} \frac{T(\hat{\mathbf{q}}_2)}{T_0} \right\rangle, \tag{7.9b}$$

and so on. These correlation functions depend on the orientation of the axes used to define the Stokes parameters. Their advantage is that they are easier to calculate than Eqs. (7.8) and are easier to measure when signal to noise is low; their disadvantage is that their physical interpretation is less easily visualized. Correlation functions are commonly characterized by the coefficients $C_l$ of an expansion in Legendre polynomials:

$$C^{TT}(\theta) = \sum_{l=2}^{\infty} \frac{(2l+1)}{4\pi} C_l^{TT} P_l(\cos\theta) \tag{7.10}$$

and likewise for the others. The $l = 1$ term, indistinguishable from the Doppler shift from proper motion with respect to the rest frame of the microwave background, is ignored. Note this conventional normalization of $C_l$ unfortunately differs by a factor of $4\pi$ from the conventional normalization of the brightness moments, Eq. (6.6).

### A. Temperature Correlation Functions

The temperature correlation function $C^{TT}(\theta)$ can be evaluated exactly in terms of the brightness moments. First, since the equations for $\Delta(\mathbf{K})$ depend only on $|\mathbf{K}|$, separate out the angular dependence by writing the fluctuations as

$$\Delta_{Il}^i(\mathbf{K}) = \tilde{\Delta}_{Il}^i(K)\zeta_i(\mathbf{K})e^{i\delta_i(\mathbf{K})}, \tag{7.11}$$

where $i$ stands for $s$, $+$, or $\times$. The functions $\zeta_i$ and $\delta_i$ are random variables set by the initial conditions: $\zeta_i$ is real and positive, normalized such that $\int d\Omega_K \zeta_i^2(\mathbf{K}) = 4\pi$, and the phase $\delta_i$ is a real number between 0 and $2\pi$. For gaussian initial conditions, $\zeta_i$ is drawn from a normal distribution and $\delta_i$ is uniformly distributed, independently for each $\mathbf{K}$ mode and each type of perturbation.

Substituting Eq. (7.1) into Eq. (7.7) provides the starting point for evaluating the temperature correlation functions. The average value can be replaced by $V^{-1}\int d^3\mathbf{x}$, where $V$ is the sample volume, and in the limit of a large volume the sums over $\mathbf{K}$ vectors can be replaced by integrals: $\sum_\mathbf{K} \to V/(2\pi)^3 \int d^3\mathbf{K}$. The $\mathbf{x}$ integral gives a delta functional which eliminates one of the $\mathbf{K}$ integrals. For scalar perturbations, the result is

$$\left\langle \frac{T(\hat{\mathbf{q}}_1)}{T_0}\frac{T(\hat{\mathbf{q}}_2)}{T_0}\right\rangle = \frac{V}{128\pi^3}\int d^3\mathbf{K}\sum_{l_1 l_2}(2l_1+1)(2l_2+1)P_{l_1}(\cos\theta_1')P_{l_2}(\cos\theta_2')\Delta_{Il_1}^{s\,*}(\mathbf{K})\Delta_{Il_2}^{s}(\mathbf{K}). \tag{7.12}$$

Then expressing each Legendre polynomial in terms of spherical harmonics using the spherical harmonic addition formula and performing the angular piece of the $\mathbf{K}$ integral using the decomposition in Eq. (7.11) gives the familiar formula

$$C_l^{TT} = \frac{V}{8\pi}\int_0^\infty K^2 dK |\tilde{\Delta}_{Il}^s(K)|^2. \tag{7.13}$$

For tensor perturbations, the added angular dependence makes the calculation somewhat more involved; in this case, Eq. (7.12) becomes

$$\left\langle \frac{T(\hat{\mathbf{q}}_1)}{T_0}\frac{T(\hat{\mathbf{q}}_2)}{T_0}\right\rangle = \frac{V}{128\pi^3}\int d^3\mathbf{K}\sum_{l_1 l_2}(2l_1+1)(2l_2+1)P_{l_1}(\cos\theta_1')P_{l_2}(\cos\theta_2')\sin^2(\theta_1')\sin^2(\theta_2')$$
$$\times\left[\cos 2\phi_1'\cos 2\phi_2'\tilde{\Delta}_{Il_1}^{+\,*}(K)\tilde{\Delta}_{Il_2}^{+}(K)\zeta_+(\mathbf{K})^2 + \sin 2\phi_1'\sin 2\phi_2'\tilde{\Delta}_{Il_1}^{\times\,*}(K)\tilde{\Delta}_{Il_2}^{\times}(K)\zeta_\times(\mathbf{K})^2\right]. \tag{7.14}$$

The cross-terms between the two types of tensor perturbations (and between the tensor and scalar perturbations) cancel because the integral over $\mathbf{K}$ contains random phase factors, e.g. $\exp(i\delta_+(\mathbf{K}) - i\delta_\times(\mathbf{K}))$. Rearranging the terms in brackets yields

$$\left\langle \frac{T(\hat{\mathbf{q}}_1)}{T_0}\frac{T(\hat{\mathbf{q}}_2)}{T_0}\right\rangle = \frac{V}{128\pi^3}\int d^3\mathbf{K}\sum_{l_1 l_2}(2l_1+1)(2l_2+1)P_{l_1}(\cos\theta_1')P_{l_2}(\cos\theta_2')\sin^2(\theta_1')\sin^2(\theta_2')$$
$$\times\frac{1}{2}\Big[\cos(2\phi_1'+2\phi_2')\left(\tilde{\Delta}_{Il_1}^{+\,*}(K)\tilde{\Delta}_{Il_2}^{+}(K)\zeta_+(\mathbf{K})^2 - \tilde{\Delta}_{Il_1}^{\times\,*}(K)\tilde{\Delta}_{Il_2}^{\times}(K)\zeta_\times(\mathbf{K})^2\right)$$
$$+\cos(2\phi_1'-2\phi_2')\left(\tilde{\Delta}_{Il_1}^{+\,*}(K)\tilde{\Delta}_{Il_2}^{+}(K)\zeta_+(\mathbf{K})^2 + \tilde{\Delta}_{Il_1}^{\times\,*}(K)\tilde{\Delta}_{Il_2}^{\times}(K)\zeta_\times(\mathbf{K})^2\right)\Big]. \tag{7.15}$$

Further progress can be made with the mild and physically reasonable assumption that $\tilde{\Delta}_{Il}^+(K) = \tilde{\Delta}_{Il}^\times(K)$, in other words that the power spectra for the two polarizations of gravity waves are the same. Then the two terms on the second line of the above equations give equal integrals and cancel. The remaining trigonometric functions can be written in a form independent of $\phi'$ [41]:

$$\left\langle \frac{T(\hat{\mathbf{q}}_1)}{T_0} \frac{T(\hat{\mathbf{q}}_2)}{T_0} \right\rangle = \frac{V}{128\pi^3} \sum_{l_1 l_2} (2l_1+1)(2l_2+1) \int dK\, K^2 \tilde{\Delta}_{I l_1}^{+*}(K) \tilde{\Delta}_{I l_2}^{+}(K)$$

$$\times \int d\Omega_{\mathbf{K}} P_{l_1}(\mu'_1) P_{l_2}(\mu'_2) \left[2(\hat{\mathbf{q}}_1 \cdot \hat{\mathbf{q}}_2 - \mu'_1 \mu'_2)^2 - (1-\mu'^2_1)(1-\mu'^2_2)\right], \quad (7.16)$$

where as in the scalar case the $\mathbf{K}$ integral has been separated into its magnitude and angular dependences, and $\mu'_1 = \cos\theta'_1$, etc. Now the evaluation and simplification of the angular integral is a straightforward but lengthy process. The various factors of $\mu$ may be absorbed into the Legendre polynomials using the recursion relation

$$(2l+1)xP_l(x) = (l+1)P_{l+1}(x) + lP_{l-1}(x). \quad (7.17)$$

The angular integral may be performed using the same procedure as in the scalar case: replace the Legendre polynomials with spherical harmonics using the addition theorem, and use the orthogonality of the spherical harmonics to eliminate the integrals. Then converting the remaining spherical harmonics back to Legendre polynomials gives many terms proportional to $P_l(\hat{\mathbf{q}}_1 \cdot \hat{\mathbf{q}}_2)$ with various indices $l$. The original factors of $\hat{\mathbf{q}}_1 \cdot \hat{\mathbf{q}}_2$ in Eq. (7.16) can now be absorbed into the Legendre polynomials using the above recursion relation. At this point, the terms can be collected together, and after much algebra the formula finally simplifies to [40]

$$C_l^{TT} = \frac{V}{8\pi} \frac{(l+2)!}{(l-2)!} \int dK\, K^2 \left| \frac{\tilde{\Delta}_{I l-2}^{+}(K)}{(2l-1)(2l+1)} - 2\frac{\tilde{\Delta}_{I l}^{+}(K)}{(2l-1)(2l+3)} + \frac{\tilde{\Delta}_{I l+2}^{+}(K)}{(2l+1)(2l+3)} \right|^2. \quad (7.18)$$

The total temperature correlation function is given by the sum of the scalar and tensor pieces as long as no correlation exists between the two perturbations, as will be the case for any inflationary scenario.

### B. Polarization Correlation Functions

Direct evaluation of correlation functions involving polarization involves further complications. For correlations involving the polarization vector $\mathbf{P}$ or its magnitude, substantial simplification is not possible: since the polarization vector is not linear in the Stokes parameters, the $\mathbf{x}$ integral (replacing the ensemble average) cannot be immediately performed as in the temperature correlation functions above, and no progress can be made in simplifying the general expressions for the correlation functions. The evaluation of Eqs. (7.8) must be performed numerically through, for example, a Monte Carlo average over random pairs of directions separated by a fixed angle at a given point in space.

However, the correlation functions of the Stokes parameters themselves, Eqs. (7.9), can be simplified if a small-angle approximation is invoked. The additional approximation is needed because, in contrast with the temperature case, the $Q$ and $U$ brightnesses have factors involving the rotation angle $\xi'$. Consider the $\langle QQ \rangle$ correlation function, first for scalar perturbations. Choose the axis of the spherical coordinate system to be one of the two observation directions. As above, the ensemble average can be replaced by a space integration, which then eliminates one of the $\mathbf{K}$ integrals, resulting in the expression

$$\left\langle \frac{Q(\hat{\mathbf{q}}_1)}{T_0} \frac{Q(\hat{\mathbf{q}}_2)}{T_0} \right\rangle = \frac{V}{128\pi^3} \int d^3\mathbf{K} \sum_{l_1 l_2} (2l_1+1)(2l_2+1) P_{l_1}(\cos\theta'_1) P_{l_2}(\cos\theta'_2) \Delta_{Q l_1}^{s*}(\mathbf{K}) \Delta_{Q l_2}^{s}(\mathbf{K}) \cos 2\xi'_1 \cos 2\xi'_2, \quad (7.19)$$

with $\hat{\mathbf{q}}_1 = \hat{\mathbf{z}}$, which implies $\theta'_1 = \theta_K$ and $\phi'_1 = -\phi_K$. Now if $\hat{\mathbf{q}}_1$ and $\hat{\mathbf{q}}_2$ point nearly in the same direction, then $\cos 2\xi'_1 \approx \cos 2\xi'_2 \approx \cos 2\phi_K$, using Eq. (7.3). Then the cosine factors in the above equation can be rewritten as $(1 + \cos 4\phi_K)/2$. The first term then looks just like the temperature case considered earlier. The second term can be somewhat simplified by using the spherical harmonic summation formula to rewrite $P_{l_2}$ and then explicitly performing the integral over $\phi_K$. The final result is

$$C_l^{QQ} \approx \frac{V}{16\pi} \left[ \int_0^\infty K^2 dK \left| \Delta_{Ql}^{s}(K) \right|^2 \right.$$

$$\left. + \frac{1}{4} \cos 4\phi \sum_{l_1 l_2} (2l_1+1)(2l_2+1) \frac{(l_2-4)!}{(l_2+4)!} A^4_{l l_2} A^4_{l_1 l_2} \int_0^\infty K^2 dK\, \Delta_{Q l_1}^{s*}(K) \Delta_{Q l_2}^{s}(K) \right], \quad (7.20)$$

$$A_{l_1 l_2}^m \equiv \int_{-1}^{1} dx P_{l_1}(x) P_{l_2}^m(x),$$

where the sum over $l_2$ begins at $l_2 = 4$. Note the final result depends explicitly on the azimuthal angle $\phi$ since the value of $Q$ changes if the axes are rotated.

For tensor perturbations, the drill is now familiar. The assumption that $\tilde{\Delta}_{Ql}^+(K) = \tilde{\Delta}_{Ql}^\times(K)$ with the above small-angle approximation gives the expression

$$\left\langle \frac{Q(\hat{\mathbf{q}}_1)}{T_0} \frac{Q(\hat{\mathbf{q}}_2)}{T_0} \right\rangle \approx \frac{V}{128\pi^3} \sum_{l_1 l_2} (2l_1 + 1)(2l_2 + 1) \int_0^\infty K^2 dK \tilde{\Delta}_{Q l_1}^{+\,*}(K) \tilde{\Delta}_{Q l_2}^{+}(K)$$
$$\times \int d\Omega_K \left[ P_{l_1}(\cos\theta_1') P_{l_2}(\cos\theta_2') \left( (1 + \cos^2\theta_1')(1 + \cos^2\theta_2') \cos^2 2\phi_K + 4\cos\theta_1'\cos\theta_2' \sin^2 2\phi_K \right) \right]. \quad (7.21)$$

Then the factors of $\cos\theta$ can be absorbed into the Legendre polynomials using the recursion relation. The final expression for tensor perturbations is

$$C_l^{QQ} \approx \frac{V}{16\pi} \left[ \int_0^\infty K^2 dK \left( |B_l^2|^2 + 4|B_l^1|^2 \right) \right.$$
$$\left. + \frac{1}{4} \cos 4\phi \sum_{l_1 l_2} \frac{(l_2 - 4)!}{(l_2 + 4)!} A_{l l_2}^4 A_{l_1 l_2}^4 \int_0^\infty K^2 dK \left( B_{l_1}^{2\,*}(K) B_{l_2}^2(K) + 4 B_{l_1}^{1\,*}(K) B_{l_2}^1(K) \right) \right], \quad (7.22)$$

$$B_l^1(K) \equiv (l+1)\tilde{\Delta}_{l+1}^+(K) + l\tilde{\Delta}_{l-1}^+(K),$$

$$B_l^2(K) \equiv \frac{(l+2)(l+1)}{2l+3} \tilde{\Delta}_{l+2}^+(K) + 2\frac{6l^3 + 9l^2 - l - 2}{(2l+3)(2l+1)} \tilde{\Delta}_l^+(K) + \frac{l(l-1)}{2l-1} \tilde{\Delta}_{l-2}^+(K).$$

The expressions for $\langle QT \rangle$ can be obtained in the same way; they are given in Ref. [43]. For gaussian initial perturbations, the polarization map can be reconstructed from the temperature map and the correlation functions $\langle QQ \rangle$ and $\langle QT \rangle$, as shown in Ref. [20]. In principle, all of the correlation functions involving the polarization vector are obtainable from these as well, although there appear to be no simple formulas connecting the two sets of correlation functions.

## VIII. DISCUSSION

The detection of microwave background polarization is very difficult. The most optimistic scenarios predict polarization no larger than 10% of the temperature fluctuations, or a few parts in a million of the temperature. Needless to say, this sensitivity is hard to attain. The same backgrounds which affect the temperature measurements will also affect polarization measurements. One advantage of measuring polarization which has been realized for a long time is that the experiment can chop between two orthogonal polarizations on the same patch of sky, which involves rotating a polarizer, instead of mechanically repointing the telescope. In practice, temperature measurements can only chop at a few Hz, while polarization measurements can chop at hundreds of Hz. Any atmospheric noise is thus supressed much more effectively. A mitigating effect is that since the orientation of the horn is important, side lobes from the ground and diffraction effects can add noise differently to the two polarization channels. As with temperature measurements, the ultimate experimental limit is astrophysical foreground sources, particularly our own galaxy.

On the theoretical side, where does polarization fit into the systematic investigation of microwave background anisotropies? In cosmological scenarios which invoke adiabatic initial perturbations, as in Cold or Mixed Dark Matter, the temperature correlation function, Eq. (7.7), contains features from which cosmological information may be extracted, provided reionization did not occur too early. If the universe evolved according to this broad class of models, measurement of the anisotropy spectrum will provide a detailed picture of the early linear growth of perturbations. In this case, the expected polarization can be calculated and its detection at the predicted level will provide important confirming evidence for the theory. Conversely, the angular scale and amplitude for the peak of the polarization correlation function, likely the easiest polarization quantity to measure, can serve as an additional piece of information when constraining theories. Also, precision calculations of the temperature anisotropies in these models should incorporate the polarization contribution to the temperature source term in Eqs. (6.7) and (6.9); the polarization may affect the power spectrum at the few percent level [42].

If the universe did not start out with adiabatic fluctuations, as in topological defect models, or underwent early reionization, polarization fluctuations may take on added importance in understanding the nature of the linear perturbations, because less information will be encoded in the temperature fluctuation spectrum. Because of computational difficulties in the cosmic string scenario, currently the most viable defect model, no detailed temperature fluctuation spectra including all relevant physical processes has yet been produced, and polarization has not been investigated. Owing to the more complex nature of the underlying physics, it is reasonable to expect that cosmological information will be more difficult to recover from the cosmic string temperature fluctuation spectrum than in the adiabatic case, and then polarization may provide useful additional information. Likewise, if the universe underwent early reionization prior to $z = 20$, much of the information in the adiabatic microwave background spectrum will be washed out. In this case, the level of polarization will be significantly larger [9], making it easier to detect, and may partially compensate for the temperature spectrum's loss of information. Additionally, in a reionization scenario, substantial secondary anisotropies in both temperature and polarization may be produced via the Ostriker-Vishniac effect [24]. Second-order effects on polarization have not yet been systematically investigated.

Polarization limits or detections may also give useful information about intergalactic magnetic fields at early times. Specifically, if a net circular polarization is ever detected and convincingly separated from foreground contamination, this would be a strong suggestion for a magnetic field polarizing the free electrons. Even upper limits on polarization provide one of the only methods of constraining the primordial magnetic field.

Finally, since temperature and polarization couple differently to tensor and scalar metric perturbations, it may be possible to separate the two contributions with a combination of temperature and polarization measurements. Initial steps in this direction have been taken in Ref. [43], which considers differences in $\langle QT \rangle$ between tensor and scalar perturbations. The overall size of the correlations are of course small, but some differences do appear in the cross-correlation function. A different correlation function or some other type of statistic optimized to look for correlations peculiar to tensor perturbations, may give more promising results.

## ACKNOWLEDGMENTS

I would like to thank Scott Dodelson for teaching me the Boltzmann equation approach to the microwave background, Gunter Sigl for patient explanations of his work, and Lloyd Knox for penetrating questions. I would also like to acknowledge helpful conversations with Stephan Meyer and Robert Crittenden. Michael Turner has continually provided support and encouragement. Thanks also to Edward Kolb, Tom Witten, and Rene Ong for their time and energy. This work was supported primarily by the NASA Graduate Student Researchers Program, and in part by the DOE (at Chicago and Fermilab) and by NASA through grant No. NAGW 2381 (at Fermilab).